\documentclass[lettersize,journal]{IEEEtran}

\usepackage{amsmath,amsfonts}
\usepackage{array}
\usepackage[caption=false,font=normalsize,labelfont=sf,textfont=sf]{subfig}
\usepackage{textcomp}
\usepackage{stfloats}
\usepackage{url}
\usepackage{verbatim}
\usepackage{graphicx}
\usepackage{balance}

\newtheorem{definition}{Definition}

\usepackage{xr}
\usepackage{graphicx} 
\usepackage{soul} 
\usepackage{booktabs}
\usepackage{url}

\usepackage[strings]{underscore}

\usepackage{hyperref}
\usepackage{amsfonts}
\usepackage{nicefrac}
\usepackage{microtype}
\usepackage{lipsum}
\usepackage{tabularx}
\usepackage{float}
\usepackage{dcolumn}
\usepackage{booktabs}
\usepackage{adjustbox}
\usepackage{amsmath}
\usepackage{algpseudocode}
\usepackage{algorithm}
\usepackage{comment}

\usepackage{multirow}
\usepackage[table,xcdraw]{xcolor}
\usepackage{marvosym}
\usepackage{stmaryrd}

\algnewcommand\Var[1]{\mbox{\itshape#1}}

\definecolor{seashell}{rgb}{1.0, 0.96, 0.93}

\begin{document}

\title{Unveiling Cognitive Constraints in Language Production: Extracting and Validating\\ the \emph{Active} Ego Network of Words}


\author{Kilian~Ollivier,
        Chiara~Boldrini,
        Andrea~Passarella,
        and~Marco~Conti
\thanks{This work was partially supported by SoBigData.it. SoBigData.it receives funding from European Union – NextGenerationEU – National Recovery and Resilience Plan (Piano Nazionale di Ripresa e Resilienza, PNRR) – Project: “SoBigData.it – Strengthening the Italian RI for Social Mining and Big Data Analytics” – Prot. IR0000013 – Avviso n. 3264 del 28/12/2021. C. Boldrini was  supported by PNRR - M4C2 - Investimento 1.4, Centro Nazionale CN00000013 - "ICSC -National Centre for HPC, Big Data and Quantum Computing" - Spoke 6, funded by the European Commission under the NextGeneration EU programme.}
\thanks{K. Ollivier, C. Boldrini, A. Passarella and M. Conti are with IIT-CNR, Via G. Moruzzi 1, 56124, Pisa, ITALY e-mail: \{k.ollivier, c.boldrini, a.passarella, m.conti\}@iit.cnr.it}
}


\maketitle

\begin{abstract}
The ``ego network of words'' model captures structural properties in language production associated with cognitive constraints. While previous research focused on the layer-based structure and its semantic properties, this paper argues that an essential element, the concept of an \emph{active network}, is missing. The \emph{active} part of the ego network of words only includes words that are regularly used by individuals, akin to the ego networks in the social domain, where the active part includes relationships regularly nurtured by individuals and hence demanding cognitive effort.
In this work, we define a methodology for extracting the active part of the ego network of words and validate it using interview transcripts and tweets. The robustness of our method to varying input data sizes and temporal stability is demonstrated. 
We also demonstrate that without the active network concept (and a tool for properly extracting the active network from data), the ``ego network of words'' model is not able to properly estimate the cognitive effort involved and it becomes vulnerable to the amount of data considered (leading to the disappearance of the layered structure in large datasets).
Our results are well-aligned with prior analyses of the ego network of words, where the limitation of the data collected led automatically (and implicitly) to approximately consider the active part of the network only. Moreover, the validation on the transcripts dataset (MediaSum) highlights the generalizability of the model across diverse domains and the ingrained cognitive constraints in language usage.
\end{abstract}

\begin{IEEEkeywords}
ego network of words, active network, cognitive constraints, language production, structural properties
\end{IEEEkeywords}

\section{Introduction}

\label{sec:intro}

\begin{figure}[t]
  \centering
  \subfloat[Social ego network]{
  \includegraphics[width=.45\linewidth]{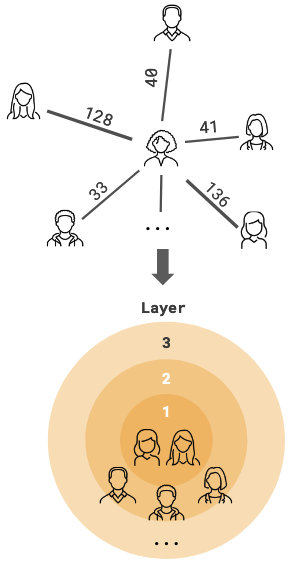} \vspace{-5pt}
  \label{fig:social-egonet}}
  \vspace{-8pt}
    \hfill
    \subfloat[Ego network of words]{
  \includegraphics[width=.48\linewidth]{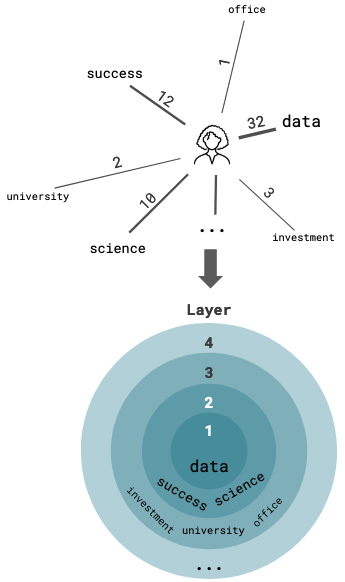}\vspace{-8pt}
  \label{fig:words-egonet}}
\vspace{5pt}
\caption{A social ego network is constructed by first calculating the frequency of contact between a person (the ego) and the individuals the ego has a social relationship with (the alters). The alters are then grouped into concentric layers according to these numbers (Figure~\ref{fig:social-egonet}). In the same way, we found that words could be grouped into concentric layers after studying their frequency of use by the ego} (Figure~\ref{fig:words-egonet}).
\label{fig:egonet-types} \vspace{-20pt}
\end{figure}

\IEEEPARstart{H}{uman} language production is subject to many cognitive processes that unfold transparently. These processes exploit our cognitive abilities (subject to physiological limits such as the duration and volume of long-term memorization of the mental lexicon) to their full extent.
For example, it is possible to find the word that best fits the idea that needs to be expressed among thousands of words in only a few milliseconds~\cite{levelt1999theory}, thanks to complex processing levels (semantic, syntactic, and lexical) involved in speech-related cognition~\cite{caramazza1997many}. The structure of the language is influenced by these cognitive strategies. For instance, in most of the still existing languages, the most frequent words of a language are both the shortest~\cite{bentz2016zipf} and the most quickly retrieved ones in a speech production task~\cite{broadbent1967word,qu2016tracking}. According to Zipf~\cite{zipf1949human}, some of these structural regularities are the result of a compromise that minimizes the effort spent in communication for both the sender -- who prefers to use frequent words to minimize the word retrieval time -- and the receiver -- who prefers less used words to minimize ambiguity. Previous work has shown the existence of a new set of structural~\cite{ollivier2020} and semantic~\cite{ollivier2022} invariants in language production using an egocentric model inspired by the social ego network model~\cite{arnaboldi2012analysis}, which in turns originates from the social brain hypothesis from anthropology~\cite{dunbar1998social}. This model organizes a person's (the ego) social relationships into concentric circles (between four and five on average) according to their intensity (a toy example is provided in Figure~\ref{fig:social-egonet}). Recent work has leveraged large amounts of data from social networks to show that this model is also relevant for describing online relationships~\cite{dunbar2015structure}.

In this paper, we adopt a similar approach to study cognitive limitations in language production. Indeed, an ego-centered model organized in concentric layers (called \emph{``ego network of words''}) can be used to describe the way a person uses his personal vocabulary (the model is illustrated in Figure~\ref{fig:words-egonet}). Language production, just like the socialization process, consumes cognitive capacities that are limited, despite the power of the human brain. These two human activities are closely connected, as postulated by the ``social gossip theory of language evolution''~\cite{dunbar1998social}, which establishes a causal link between the sudden increase in the number of active relationships in humans (from 50 for the closest non-verbal primates to 150 for humans) and the appearance of language that would have optimized the activity of social grooming. Moreover, we expected to find traces of cognitive limits in ego networks of words since we already have evidence of such limits in language production, like the size of the vocabulary, which would be about 42,000 words for a 20-year-old native English speaker, or the approximate time span of 180 ms to retrieve a word which is a strong constraint~\cite{costa2009time}.

\vspace{-10pt}
\subsection{Contributions and key results of the paper}
\vspace{-3pt}

The ego network of words is a novel model that captures structural properties in language production linked to cognitive constraints. Existing works focused on the layer-based structure and its semantic properties. Here, we argue that the model is still missing a key element used in the characterization of social ego networks, i.e., the concept of the active network. In social ego networks, the active part of the ego network only included relationships that the ego spent time nurturing, thus consuming cognitive resources on the ego's side. The layered structure of the social ego network only emerged in the active part. Such ``meaningful'' relationships were identified with a traditional anthropological approach, grounded in a shared understanding of how human social interactions work. Specifically, a relationship was considered meaningful if it entailed at least one interaction per year, based on the fact that people close to each other exchange at least birthday or holiday wishes. \footnote{These considerations hold for Western societies, which were the focus of these anthropological studies.}.

In previous works on language ego networks, the layered structure seemed to emerge without applying any preliminary filter in the spirit of the birthday/holiday wishes. And anyway, finding such a common sense threshold for the ego network of words would not have been possible. In this paper, we argue that without the notion of ``active'' ego network of words, the analysis carried out would not be robust to the amount of data considered. Specifically, in the paper we show three key properties in this regard. First, that depending on the size and extent of collected data, ego network may or may not include (a part of) the inactive ego network (i.e., the ego network beyond the active part). Second, that appropriate filtering is needed, in order to isolate the active part of the ego network. Third, that layered structures -- the fingerprint of the human cognitive involvement -- emerge only when the inactive part of the ego network is excluded. Therefore, the paper provides evidence about the complete structure of the ego network of words, as well as a robust methodology to isolate and study it.

The first contribution of this paper is the definition of a methodology to extract the ``active'' part of the ego network (Section~\ref{sec:part1}). In Section~\ref{sec:optimal-circles-active}, we successfully test this methodology using two types of datasets: interview transcripts and tweets. MediaSum is a dataset that includes thousands of verbatim transcripts of spoken interviews from an American public radio and private TV channel (Section~\ref{sec:mediasum}). The Twitter datasets are extracted from the same users as in \cite{ollivier2020}, but we downloaded larger timelines, up to 10K tweets (Section~\ref{sec:twitter}). We also prove that the method that we use to extract the active ego network is robust to different amounts of input data (Section \ref{sec:robustness}) and that the active size is stable over time (Section \ref{sec:temporal-stability}). The structural results (Section~\ref{sec:revisiting-active-egonets}) of the ego networks produced in this way substantially confirm the layer ego network of word structure obtained in previous work~\cite{ollivier2020} but are robust to the size of the input data. 
The second contribution of the paper is the validation of the ego network of words model on a dataset (MediaSum) that is completely different in nature from the Twitter ones on which it had been applied previously (Section~\ref{sec:revisiting-active-egonets}). The fact that the structural properties of the word ego networks are confirmed is an important validation that the model generalizes across different domains and, thus, that the underlying cognitive constraints are ingrained in our use of language. 

The key findings of the article are as follows: 

\begin{itemize} 
    \item We introduce the notion of \emph{active part of an ego network of words}, beyond which the model would contain words that are not used frequently enough to denote a cognitive involvement. We show that, beyond the active part, the word ego network becomes poorly structured (\textit{i.e.} with a very low number of concentric circles).
    \item We define a robust algorithm to extract this active part based on the properties of the ego's language production.
    \item We find that the active size is specific to each ego network and stable over time. Therefore, each ego appears to have its own limit to the number of words it can actively use, similarly to what was observed for social ego networks.
    \item Even if the ego networks are larger than those observed in previous papers~\cite{ollivier2020,ollivier2022} (where the concept of active network was not exploited) we retrieve most of the structural invariants previously observed: first, the number of circles in the model is approximately the same. Second, third-to-last and second-to-last circles account for 30\% and 60\% of the words in the ego network whatever the number of layers. Third, the scaling ratio between circles tends towards 2.
    \item Ego networks based on oral language production (interviews) have the same structural properties as those obtained from Tweets, thus confirming the cross-domain generalizability of the ego network model.
\end{itemize}

\noindent
To the best of our knowledge, our work is the first one attempting to define a methodology for the extraction of the active part of ego networks, both in the social and in the language domain.

\vspace{-15pt}
\section{Related work}
\vspace{-3pt}

\subsection{Social ego networks}

The social ego network model organizes the interpersonal relationships of a person (the ego) into concentric circles. This is an empirical model derived from the work of anthropologist Robin Dunbar on the number of active relationships that a human can maintain on average over time~\cite{dunbar1998social}.  To do this, he established a correlation between the relative size of a part of the brain dedicated to sociability (the neocortex) and the typical group size in primates, then deduced what the equivalent number would be for humans. This number, 150, is called Dunbar's number. Anthropological studies have shown that this number is a recurrent occurrence in human organizations, as can be observed in Hutterite communities where it is the maximum number before the group splits up, in Israeli Kibbutzim where it is the average number at the foundation time, but also in modern factories sizes~\cite{Dunbar.2018}. By analyzing the traces left by online social interactions, researchers have shown that the number of active online relationships that can be maintained at the same time is in the same order of magnitude as Dunbar's number~\cite{dunbar2015structure}. Moreover, for a given person (the ego) it is possible to subdivide these active relationships (alters) into four concentric circles~\cite{hill2003social,Zhou2005}, the most central one containing the most intimate relationships. These circles contain about 5, 15, 50, 150 alters, and exhibit a consistent scaling ratio of three in their sizes. This model of concentric circles, called ``ego network model'' was also confirmed for online relationships, with approximately the same number of circles and the same scaling ratio~\cite{dunbar2015structure} as for offline relations. Thanks to online social networks, we also know that after an initial moment of growth, the ego network structure remains stable over time for the majority of the individuals~\cite{arnaboldi2013dynamics,arnaboldi2014information}.

\vspace{-10pt}
\subsection{Structural and semantic properties of ego network of words}
\vspace{-3pt}

Previous papers have shown the relevance of using an ego network of words for studying language production~\cite{ollivier2020,ollivier2022}. Using datasets extracted from Twitter, ego networks of words were constructed with a methodology similar to that used to construct social ego networks. However, instead of considering other people as alters and the frequency of contact with the ego as a proxy for the intensity of the relationship, words were considered as alters, and their proximity to the ego is measured by their frequency of use. In this way, each ego's vocabulary is organized into concentric layers, the first of which would contain the most frequently used words while the last would contain the least used words. Even if, unlike social ego networks, the size of the ego network varies significantly, the number of layers remains in the same order of magnitude: between five and seven~\cite{ollivier2020}. A very strong similarity in the relative size of concentric layers between egos with the same amount of layers was found, regardless of the dataset. Moreover, the third-to-last and second-to-last layers account for 30\% and 60\% of the words in the ego network whatever the number of layers,
which means that the total number of layers depends on the number of internal layers (from the innermost to the third to last), which is determined by the distribution of the most frequent words. Finally, it appeared that the scaling ratio is not three as in the case of the social ego network, but tends towards two consistently when moving towards the outer layers. A semantic analysis of the rings was also performed, assigning each one a semantic identity card~\cite{ollivier2022}. This is a distribution of the importance given to one hundred topics found automatically and common to a whole dataset. We found that the innermost ring is the most different from the others, as it generates proportionally more topics. All the important topics of this ring are also important in the whole ego network and vice versa. That is why this layer can be seen as the semantic fingerprint of the ego network.



\vspace{-15pt}
\section{The datasets}
\label{sec:dataset}


In this study, we will rely on two types of datasets. The first, MediaSum~\cite{zhu2021mediasum}, compiles years of television and radio interview transcripts. In the second, we collected up to ten thousand tweets each from four distinct groups of Twitter users.

\vspace{-15pt}
\subsection{MediaSum \label{sec:mediasum}}

MediaSum contains about 464K interview transcripts, of which 49K are from NPR (American public radio) and 415K from CNN (cable news channel). These interviews are extracted from well-known broadcasts, such as ``Anderson Cooper 360 degrees'' on CNN or ``Morning Edition'' on NPR. This is a valuable dataset, as it allows us to study the ego networks of words produced from spoken-language corpora collected over a long period of time. Indeed, the dataset contains between 10K and 35K interviews per year between 2000 and 2020. 
The speakers are mainly television or radio anchors and recurring guests. Another advantage is that the topics of the interviews are diverse (\textit{eg.} politics, international news, crime), and so are 
the guests such as the athlete Michael Phelps or the actor Morgan Freeman. Each interview lasts, on average, 30 turns (each turn corresponds to a speaker's line of dialogue that we call ``utterance'') and involves 6.5 speakers (4.0 for NPR and 6.8 for CNN). 
Taking into account its characteristics, this dataset is particularly interesting for investigating the long-term cognitive limitations related to the language of various kinds of people.


\subsubsection{Cleaning the dataset\label{sec:MediaSum-clean}}
\begin{figure}[!t]
  \centering
  \includegraphics[width=0.6\linewidth]{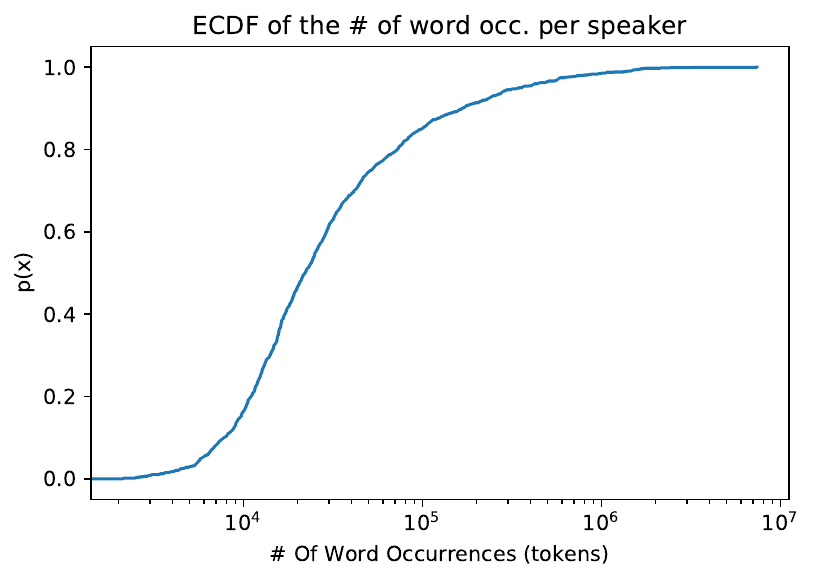}\vspace{-8pt}
  \caption{Word occurrences per speaker.}
  \label{fig:tok-per-speaker}
\end{figure}

\begin{figure}[!t]
  \centering
  \includegraphics[width=0.6\linewidth]{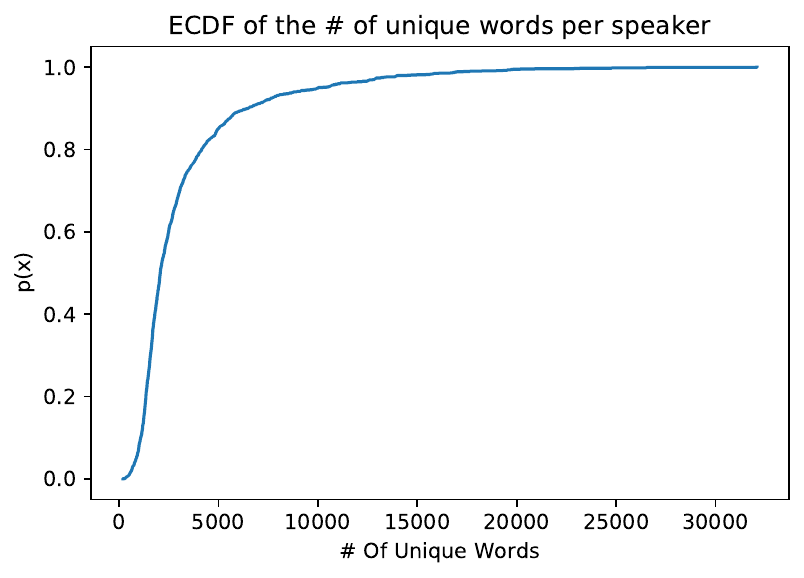}\vspace{-8pt}
  \caption{Unique words per speaker.}\vspace{-10pt}
  \label{fig:word-per-speaker}
\end{figure}

~\\
Since we want to group all of the dialogue lines for each person across the entire dataset, we must first clean the names which are manually filled (\textit{eg.} ``wozniak'', ``steve wozniak'', ``steve wozniak, founder, apple computer'', ``mr. steve wozniak (co-founder, apple computer)''). After this name-cleaning operation and a first round of deletion of speakers with too few utterances (mainly due to inconsistencies in their names like spelling mistakes), we end up with 106,627 speakers. The average number of utterances per speaker is around 124 (Table~\ref{tab:cleaning-figures}). In our previous papers \cite{ollivier2020,ollivier2022}, where we used corpora extracted from Twitter, we defined a minimum of 500 tweets per user. In a similar way, we keep only speakers with at least 500 utterances such that the corpora to process have a minimum size. This criterion results in the suppression of 98.6\% of the speakers, but only 55\% of the total number of utterances in the dataset. This relatively small group of speakers produces almost half of the text corpus that we will use to build ego networks of words. 
The sentences are tokenized, the stop words are removed and the remaining tokens are lemmatized to group together inflected versions of the same word. Once we obtain the number of words' occurrences for a given speaker, we remove those that appear only once to leave out most misspelled words.
As we can see in Figure~\ref{fig:tok-per-speaker} and Figure~\ref{fig:word-per-speaker}, a few speakers have a very large number of word occurrences and unique words. Unsurprisingly, most of them are anchormen or anchorwomen, like Wolf Blitzer of CNN, who are the most active speakers in the dataset. The majority of speakers have between 10K and 100K word occurrences and less than 5K unique words. The average number of word occurrences among all the speakers is 89,313 and the average number of unique words is 5,316.

\begin{table}[t!]
\begin{center}
\caption{MediaSum statistics, before and after removing speakers with less than 500 utterances (word stats are only computed for users with $>500$ utterances).}
\label{tab:cleaning-figures}
\begin{tabular}{@{}lrr@{}}
\toprule
 & \textbf{Before} & \textbf{After} \\
\midrule
Number of speakers & $106,627$ & $1,513$ \\
Number of utterances & $13,228,854$ & $5,931,363$ \\
Number of utterances / speaker & $124$ & $3,920$ \\
Number of words / speaker & $-$ & $89,313$ \\
Number of unique words / speaker & $-$ & $5,316$ \\
\bottomrule
\end{tabular}
\end{center}
\end{table}



\vspace{-10pt}
\subsection{Twitter\label{sec:twitter}}

In \cite{ollivier2020,ollivier2022}, we built ego networks of words based on Twitter timelines with up to 3.2K tweets (the download limitation of the standard Twitter API) collected from four sets of users:

\begin{LaTeXdescription}
\item[Journalists] working for the New York Times. The NYT itself has created a list of 678 accounts \footnote{\url{https://twitter.com/i/lists/54340435}}.
\item[Science writers] who tweet about science-related topics. A list of 497 accounts has been created by Jennifer Frazer\footnote{\url{https://twitter.com/i/lists/52528869}}, a science writer at \textit{Scientific American}.
\item[Random users \#1] are sampled among accounts that published on January 16, 2020 (download time) a tweet or a retweet in English containing the hashtag \textit{\#MondayMotivation}. This hashtag, which is both popular and neutral, does not refer to a political or controversial issue. Bot accounts are filtered using the Botometer service~\cite{davis2016botornot}, which leverages both structural properties (number of followers, tweeting frequency, \textit{etc}) and language features to detect non-human behaviors. After this operation, the dataset contains 5,183 accounts.
\item[Random users \#2] are sampled among accounts that issued on February 11\textsuperscript{th} 2020 (download date) a tweet or a retweet in English, from the United Kingdom. The group contains 2,733 accounts after removing the bots.
\end{LaTeXdescription}

\begin{figure}[t]
    \begin{center}
        \includegraphics[width=\linewidth]{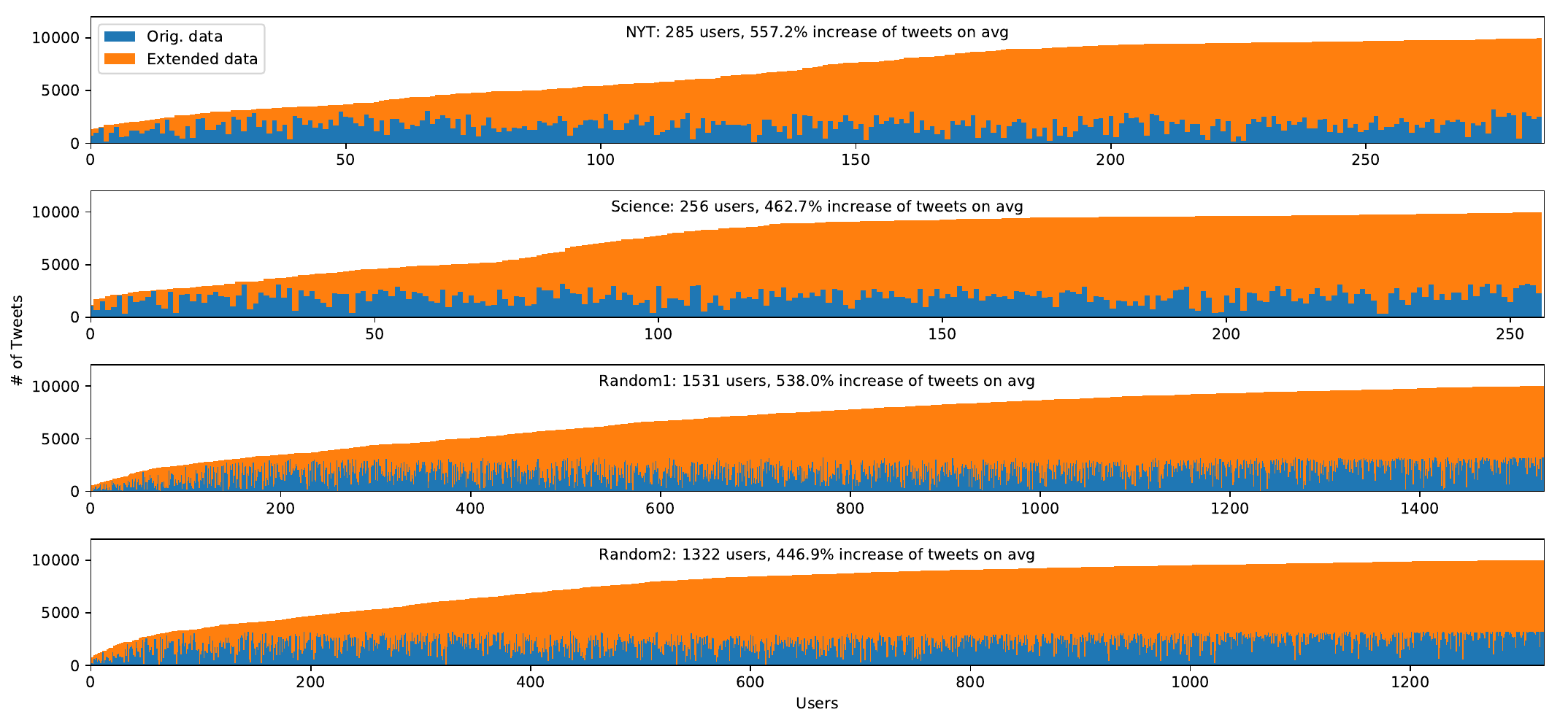}
        \caption{Collected Twitter timelines containing at least 500 tweets. Each bar corresponds to a timeline, where the blue part refers to the number of tweets in the original dataset, and the orange part refers to the number of newly collected tweets.}
        \label{fig:twitter-extension}
    \end{center}
\end{figure}

\noindent
We extended the timelines of these four sets of users to up to 10K tweets, by leveraging the extended download capabilities of the Twitter Academic Research track. As illustrated by Figure~\ref{fig:twitter-extension}, this results in much longer timelines with respect to those analysed in previous works. These longer timelines are used to stress-test the ego network of words model.
In the same fashion as in \cite{ollivier2020,ollivier2022}, and in Section \ref{sec:MediaSum-clean}, we only keep the timelines with at least 500 tweets. The figures related to the number of word occurrences and unique words are reported in Table~\ref{tab:twitter-figures}. Even if the numbers are lower for both random user datasets compared to journalists and science writers, all figures are of the same order of magnitude as for MediaSum.

\begin{table}[!t]
\begin{center}
\caption{Twitter datasets after removing users with $<$ 500 tweets.}
\label{tab:twitter-figures}
\setlength{\tabcolsep}{0.5em}
\begin{tabular}{@{}lrrrr@{}}
\toprule
\textbf{Dataset} & \textbf{\# of users} & \textbf{Word occ./user} & \textbf{Unique words/user}\\
\midrule
NYT journalists & $285$ & $87,698$ & $11,877$ \\
Science Writers & $256$ & $138,050$ & $14,952$ \\
Random users \#1 & $1,536$ & $48,021$ & $6,650$ \\
Random users \#2 & $1,324$ & $57,177$ & $6,757$ \\
\bottomrule
\end{tabular}
\end{center}
\end{table}

\vspace{-10pt}
\section{Methodology}
\label{sec:part1}

\vspace{-5pt}
\subsection{Preliminaries}
\label{sec:preliminaries}

Before describing our method for building the ego network of words and extracting its active part, we introduce here the notation used in the section (also summarised in Table~\ref{tab:notation}). 
We denote an ego with the letter $e$, where the ego is the speaker (MediaSum) or user (Twitter) in our datasets for whom we want to extract the ego network of words. After the cleaning process discussed in Section~\ref{sec:dataset}, for each ego $e$ we end up with a tuple (i.e., an ordered sequence) of tokens~\cite{mogotsi2010christopher}, which we denote with $\mathcal{T}_e$. Note that the tokens in $\mathcal{T}_e$ are generally not unique. In computational linguistics, the term \emph{type} denotes the class of all tokens containing the same character sequence~\cite{mogotsi2010christopher}. In other words, the set of types corresponds to the set of distinct tokens or, slightly simplifying, a type is a word and its occurrences are tokens. For example, in the sentence \emph{a rose is a rose is a rose}, there are eight tokens but only three types. In this paper, for the sake of simplicity, we use the terms \emph{type} and \emph{word} interchangeably. Similarly, tokens may be also called \emph{occurrences}. In the following, we denote the tuple of unique words in an ego network as $\mathcal{W}_e$. Please note that both $\mathcal{T}_e$ and $\mathcal{W}_e$ are ordered sequences, where the order is defined by the appearance in the ego's timeline in chronological order. 
So, if we observe the first $n$ tokens in the ego's timeline, we will get exactly $n$ tokens but at most $n$ unique words. We denote with $\mathcal{T}_e^n$ and $\mathcal{W}_e^n$ the tuples of tokens and unique words, respectively, observed up to $n$. We call $n_f$ the maximum value of $n$ (corresponding to the overall number of tokens in the observed timeline for ego $e$) such that $\mathcal{T}_e^{n_f} = \mathcal{T}_e$ and $\mathcal{W}_e^{n_f} = \mathcal{W}_e$, where $|\mathcal{T}_e| = n_f$ and $|\mathcal{W}_e| \leq n_f$.

In the rest of the section, when there is no risk of ambiguity, we will drop the subscript $e$ from our notation: so, all the variables discussed will be referring to the same tagged ego $e$.

\begin{table}[!t]
\begin{center}
\caption{Summary of notation used in the paper.}
\label{tab:notation}
\begin{tabular}{@{}ll@{}}
\toprule
\textbf{Symbol}     & \textbf{Description}         \\ \midrule
$\mathcal{T}_e$     & tuple of tokens, i.e., sequence of words ego $e$ has used  \\
$\mathcal{W}_e$     & tuple of unique words used by ego $e$        \\
$\mathcal{T}_e^n$   & $\mathcal{T}_e$ cut at the $n$-th token   \\
n & length of  the tuple $\mathcal{T}_e^n$ \\
$\mathcal{W}_e^n$   & unique words in $\mathcal{T}_e^n$\\
$w_e^n$ & length of the tuple $\mathcal{W}_e^n$ \\
$n_f$ & overall number of tokens in the observed timeline for ego $e$\\
$n_a$ & active network cut-off\\
$\tau_e$            & optimal number of circles           \\
$r_i$                & $i$-th ring of the ego network        \\
$l_i$                & $i$-th layer of the ego network       \\
$\mathcal{W}_{e,r_i}$ & unique words assigned to ring $r_i$ \\
\bottomrule \vspace{-15pt}
\end{tabular}
\end{center}\vspace{-15pt}
\end{table}

\vspace{-10pt}
\subsection{Legacy method for building an ego network of words}
\label{sec:building-egonet}
\vspace{-3pt}

Ego networks of words are used to hierarchise the words used by a given person based on their frequency. In the following, we summarise the model presented in~\cite{ollivier2022}. Let us focus on a tagged ego $e$ (hence, hereafter we drop the subscript $e$ in the notation). The ego network of words model is such that each word from $\mathcal{W}$ is assigned to one of $\tau$ rings $r_1$, $r_2$, $\ldots$, $r_{\tau}$, knowing that $r_1$ (the innermost ring) contains the most frequently used words and that $r_{\tau}$ (the outermost ring) contains the least used words. The set of words assigned to the ring $r_i$ is called $\mathcal{W}_{r_i}$ such that:

\vspace{-5pt}
\begin{equation}
        \mathcal{W} = \bigcup_{i=1}^{\tau} \mathcal{W}_{r_i}. 
    \label{eq:stability-eq1}
\end{equation}

The ego network can also be studied from a cumulative perspective with concentric layers $l_1$, $l_2$, $\ldots$, $l_{\tau}$, with layer $l_i$ containing all the rings $r_j$ where $j \leq i$.  The set of words assigned to layer $l_i$ is denoted with $\mathcal{W}_{l_i}$, so:
%
    $\mathcal{W}_{l_i} = \bigcup_{j=1}^{i} \mathcal{W}_{r_j}$.
%
This implies that the innermost layer $l_1$ is equivalent to $r_1$.
Based on these remarks, a formal definition of the problem addressed in the paper is the following.

\begin{definition}
For each observed user, given a timeline of user activity corresponding to an observed number of tokens $n_f$, obtain its ego network of words characterised by:
\begin{enumerate}
    \item an active ego network cutoff $n_a$;
    \item an optimal number of circles ($\tau_e$), each grouping together words on which the user allocates similar cognitive effort;
    \item the set of unique words ($\mathcal{W}_{e,r_i}$) assigned to each of the rings $r_i$ in this structure.
\end{enumerate}
\end{definition}

Words in an ego network are characterized by their usage frequency, which corresponds to their number of occurrences divided by the observation window (which is the same for all words uttered by the same ego). To find the best natural grouping of words (i.e., to find $\tau$) we use the Mean Shift~\cite{fukunaga1975estimation} algorithm, which is able, in contrast to Jenks~\cite{jenks1977optimal} or K-Means~\cite{macqueen1967some}, to automatically optimize $\tau$, the number of groups to be found. Clustering on a unidimensional variable is equivalent to dividing the word frequencies into mutually exclusive intervals.
The Mean Shift algorithm detects clusters that correspond to the local maxima of an estimated density function of word frequencies. This function is obtained with the kernel density estimation for which the sensibility is set with a fixed parameter called the bandwidth. We apply a preliminary log-transformation to frequencies in order to compress high values and ensure that the same bandwidth setting allows peak detection for both the high- and low-frequency parts of the distribution\footnote{Note that applying a log-transformation to word frequencies is common in psychological research when studying the associated cognitive processes~\cite{brysbaert2018word}.}.
The obtained clusters of words correspond to the $\tau$ rings of the newly built ego network of words for ego $e$, $r_1$ being the cluster containing the most frequent words and $r_{\tau}$ the one containing the least frequent words.

\vspace{-15pt}
\subsection{Why we need an active ego network extraction method}
\label{sec:full-egonet}

\begin{figure}[t]
\centering
  \centering
  \includegraphics[width=.8\linewidth]{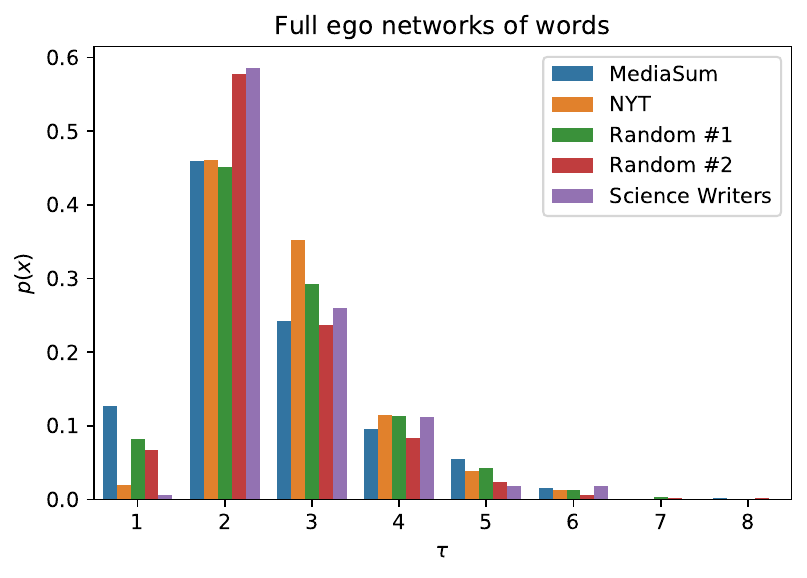}\vspace{-8pt}
  \caption{Distribution of the number of circles $\tau$ when considering all the words available in $\mathcal{W}$.}
  \label{fig:nb-circles-full} \vspace{-10pt}
  \end{figure}

\begin{figure}
  \centering
  \includegraphics[width=.8\linewidth]{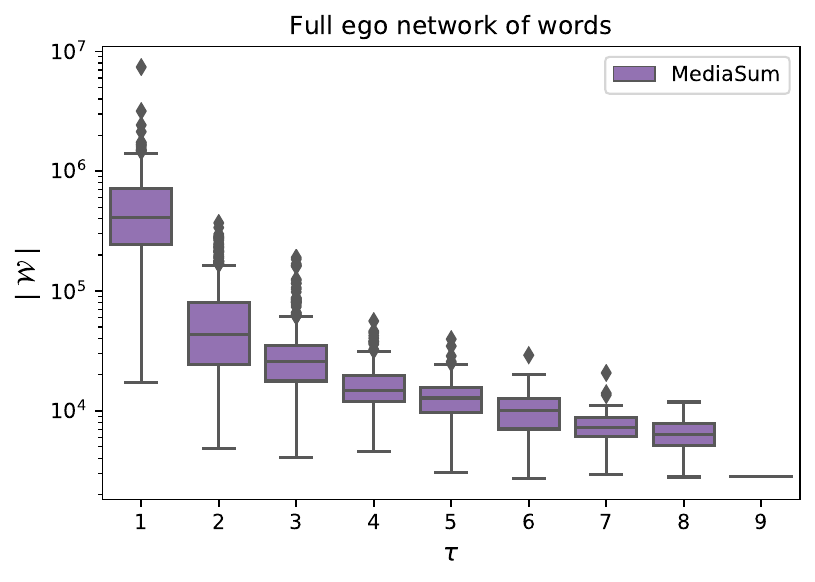}\vspace{-8pt} 
  \caption{Number of circles $\tau$ vs full ego network size ($|\,\mathcal{W}\,|$) in the MediaSum dataset. The same trend is observed in the other datasets (plots omitted to optimize the space).}
  \label{fig:occ-vs-layers} \vspace{-15pt}
\end{figure}

We start by applying the methodology described above to all the words in $\mathcal{W}$ for the egos in our datasets, and we plot the distribution of the number of circles $\tau$ in Figure~\ref{fig:nb-circles-full}. We can observe that the obtained ego networks of words have a very low number of circles (the most frequent case is two) compared with the ego networks of words in previous work (usually between five and seven circles~\cite{ollivier2020,ollivier2022}), despite exactly the same workflow being used. Note also that the Twitter datasets used here are \emph{the same} as those in ~\cite{ollivier2020,ollivier2022} except for the timeline length considered (much larger, in this work). As we can observe in Figure~\ref{fig:occ-vs-layers}, ego networks with one or two circles are the biggest ego networks (\textit{i.e.} with the largest number of unique words $|\mathcal{W}|$). This seems to suggest that, when considering larger textual inputs, the ego network model loses its finer discriminative power. In fact, two-circle ego networks are considered uninteresting, as they simply separate the most used words from the least used words. 

However, this finding is not unexpected: in the social ego network case, the theory distinguishes between the \emph{full} and \emph{active} ego network, stating that only the relationships in the active part are actually consuming cognitive resources~\cite{arnaboldi2012analysis}.
The conventional cut-off point, as stated in \cite{dunbar1998social}, is for the social relationship to involve interactions at least once a year, which, in Western societies, corresponds to at least exchanging Christmas/birthday wishes. While this cut-off point could be obtained with anthropological common sense for social ego networks, it is difficult to come up with a similar rule of thumb for the ego networks of words, which are less rooted in everyday experiences. Hence, in this work, we set out to design a methodology to automatically extract the cut-off point in the ego networks of words. This methodology should then be applied before building the ego networks as described in Section~\ref{sec:building-egonet}, in order to discard the words that do not take up cognitive capacity.

%

\vspace{-10pt}
\subsection{Extracting the active ego network}
\label{sec:active-egonet}

\begin{figure}[t]
  \centering
  \subfloat[Non-linear]{
  \includegraphics[width=.7\linewidth]{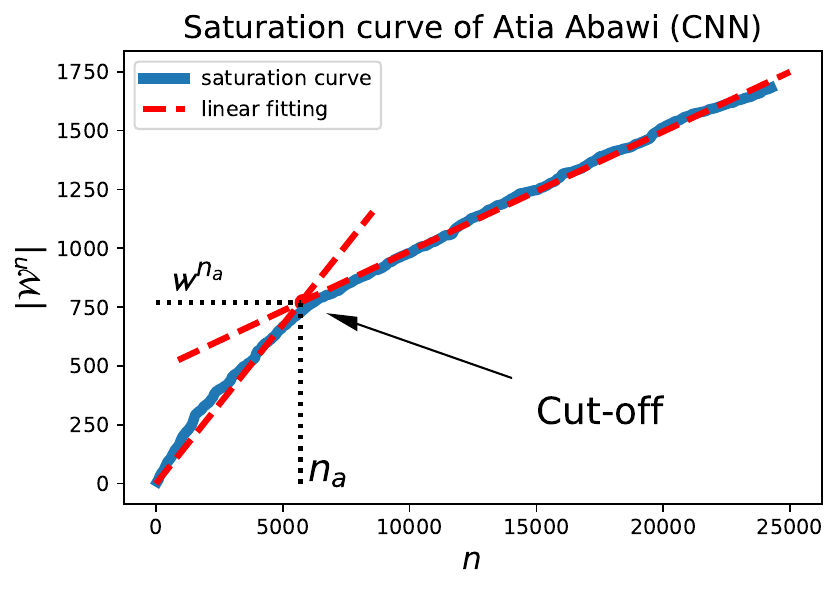} \vspace{-5pt}
  \label{fig:saturation-curve1}}
  \vspace{-8pt}
    \hfill
    \subfloat[Linear]{
  \includegraphics[width=.7\linewidth]{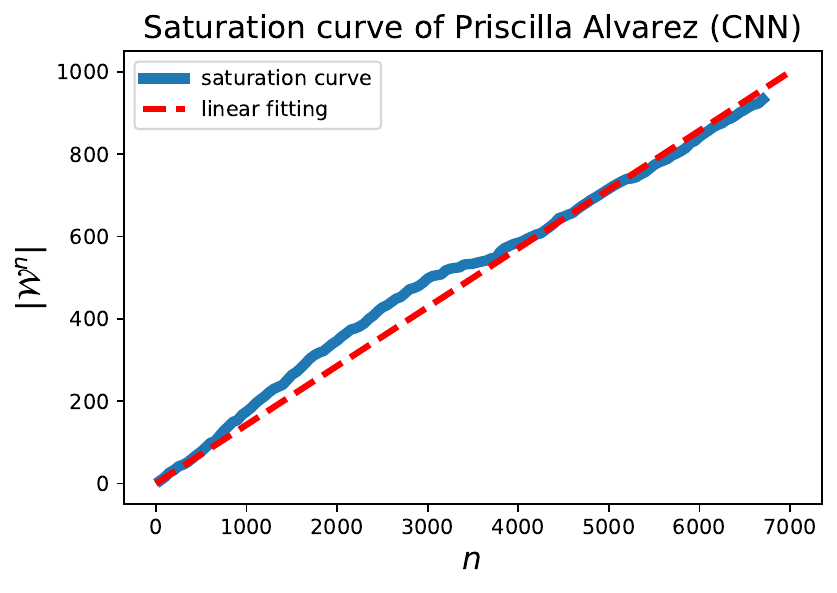}\vspace{-8pt}
  \label{fig:saturation-curve2}}
\caption{Examples of saturation curve.}
\label{fig:saturation-curve-examples}
\vspace{-10pt}
\end{figure}

The idea behind an \emph{active} ego network is that all the words it contains should be actively used, even those in the outermost circle. If we let a person speak, we notice that from a certain point on the frequency of appearance of a new word decreases rapidly:
a specific number of words is sufficient for this person to express him/herself. This quantity is the maximum number of actively used words. We can observe this phenomenon in Figure~\ref{fig:saturation-curve1}, 
where the number of tokens $n = |\mathcal{T}^n|$ vs the corresponding number of unique words $w^n = |\mathcal{W}^n|$ is plotted for a single speaker in the Mediasum dataset (we define $w^n$ to improve the readability of the formulas in the following sections). 
The curve is obtained by scanning the timeline (or, more exactly, the chronologically ordered tokens remaining after preprocessing the timeline) from start to end, and counting the tokens and the unique words as we go. The catch is that not every new token corresponds to a new unique word. 
We will call this curve the \emph{saturation curve}, which we denote with $s$. Using the notation in Section~\ref{sec:preliminaries}, $s : n \mapsto w^n$.

In Figure~\ref{fig:saturation-curve1} and~\ref{fig:saturation-curve2}, we present two typical cases observed in our datasets. Figure~\ref{fig:saturation-curve1} serves as a representative example of a broad trend that emerges in our data for users who have been observed over an extended period. Initially, there is a swift growth in the number of discovered words as new tokens are explored, but in the second phase, this growth rate significantly decreases. The rate at which new words are discovered remains fairly constant in both phases. Figure~\ref{fig:saturation-curve2} is representative of users who were not observed for a sufficient duration to reach the second phase described in Figure~\ref{fig:saturation-curve1}. In this example, the total number of tokens is much lower, comparable to the number of tokens in the initial phase for users represented in Figure~\ref{fig:saturation-curve1}.
%
We conjecture that the active part of the ego network ends at the cut-off point of the saturation curve, i.e., where the first regime ends and the second one begins. 
The saturation curve shows how many tokens are needed to observe a certain number of unique words. The number of tokens needed to increase the number of words by one can thus be seen as the maximum number of tokens an ego can use without including a new word in his spoken or written expressions. Saturation curves of ``mature'' ego networks show two regimes, whereby \emph{in the first one} words appear “sooner”, meaning that the user is able to ``resist'' less before ``injecting'' a new word.
Before proceeding further, it is important to acknowledge that in general, non-linear saturation curves may exhibit less regularity than the one depicted in Figure~\ref{fig:saturation-curve1}, while the overarching pattern of two distinct major regimes remains consistent. This might challenge algorithms intended to automatically identify the transition point between regimes. This is the rationale behind our proposal, outlined in Section~\ref{sec:methodology}, for a recursive algorithm that only terminates when the major trends are identified.

Recalling that the saturation curve is defined as $s : n \mapsto w^n$, the goal of this section is to describe a methodology for finding the value of $n$ (which we call $\hat{n}_a$) where the first phase described above ends and the second one begins. The number of unique words at the cut-off point $n_a$ of the curve corresponds to $w^{n_a} = |\mathcal{W}^{n_a}|$, while $w^{n_f} = |\mathcal{W}^{n_f}|$ corresponds to the total number of unique words in the \emph{full} ego network ($n_f$ being the maximum value of $n$). If our intuition is confirmed, the well-known layered ego network structure would emerge by considering only words in the first regime of the saturation curve when computing the ego network. Indeed, we show this in Section~\ref{sec:optimal-circles-active}.
Note that sometimes the textual data for one ego is not large enough for the ego network to reach any cut-off point (Figure~\ref{fig:saturation-curve2}). This means that the cognitive capacity for language production is not fully exploited (in the textual information available in our datasets), so the ego network of words is not fully formed. In this case, we remove the egos from the analysis because only mature ego networks are reliable for extracting structural properties.


\subsubsection{Methodology for identifying the cut-off point.}
\label{sec:methodology}

\begin{figure}[!t]
    \begin{center}
        \includegraphics[width=.9\linewidth]{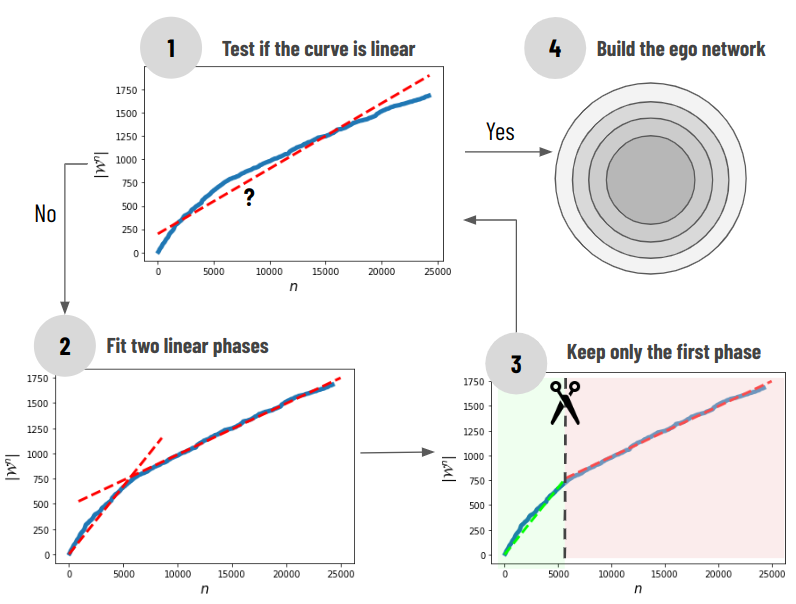}\vspace{-10pt}
        \caption{Steps for detecting the saturation point. 1) Linearity test. 2) If the curve is not linear, we find the best model fit with two linear parts. 3) We keep only the first part and repeat the process.}
        \label{fig:methodology-recurs} \vspace{-15pt}
    \end{center}
\end{figure}

We start with a high-level description of our methodology, illustrated in Figure~\ref{fig:methodology-recurs}. Let us focus on the curve~$s$, and assume that it is not linear in $[0, |\mathcal{T}|]$ (if it is linear, we can stop searching for the cut-off, since there is none). Our cut-off point $n_a$ would split~$s$ in two halves: in the first one, $s$ is approximately linear and with a greater slope; after $n_a$ the saturation curve enters a regime of reduced growth (in this second regime, $s$ might be linear or not). We want to find the knee point in $s$ where the slope change is observed. The search for $n_a$ is done recursively, continuing to split the first half until it is effectively linear. At this point, the algorithm stops. The intuition is that the words and tokens before $n_a$ correspond to the first regime described above, where new words are discovered at a higher rate. This recursive approach allows us to discard minor irregularities in the saturation curve and to properly detect the major trend of linear growth. 

Algorithm~\ref{alg:cut-off} summarises our approach. The recursive search is carried out through the \textsc{RecursiveCutOff} function, which is initially fed all data points from the saturation curve. If the saturation curve is already linear, then the algorithm returns $n_f$, the upper bound of $n$. If the saturation curve is not already linear, we need to split it into two halves. We do this with the \textsc{SplitSaturationCurve} function, which tests all the possible cut-off points and selects the one guaranteeing the best (in terms of residual sum of squares) linear fit on both sides of the cut-off. Then, we focus on the linearity of the first half to ensure there is no more potential cut-off  (we are not directly concerned with the linearity of the second part, because, as long as we are able to detect a phase change, the second part will be dropped anyway being it outside of the active network). 
What we want to assess is whether the ``signal'' in the first part of the saturation curve (before the current cut-off) is \emph{mostly} linear. To this aim, we leverage Lasso regression~\cite{tibshirani1996regression} for its ability to operate a variable reduction on its input features. The features used by Lasso are the polynomial terms of the inverse saturation curve (we consider the inverse for ease of explanation). Specifically, we consider the following: $s^{-1}(t) \sim \sum_{i=1, \ldots, p} \beta_i w^i$, with $\beta_i$ being the coefficient optimized by Lasso and $s^{-1}(t)$ the inverse of the saturation curve. In other words, we consider the growth of the number of unique words with respect to the number of tokens, and evaluate whether the dependency is mostly linear, mostly quadratic, etc. Intuitively, in the first regime of the saturation curve, the growth is linear because each new token roughly corresponds to a new unique word. Vice versa, in the second regime, we observe an inflection.  
Then, with the \textsc{LassoMaxVariableReduction} function, we denote a Lasso regression where the $\lambda$ parameter for regularization is chosen such that only one coefficient of the regression is set to a non-zero value: the one corresponding to the most significant polynomial term. If the nonzero coefficient corresponds to the linear term, we confirm that the saturation curve before the current cut-off point is linear enough for our purposes, and we stop the search. Once we obtain $n_a$, we can use it to obtain the active ego network. 
Specifically, the words in the active ego network of $e$ are  $\mathcal{W}_e^{n_a}$.

\begin{figure}[!t]
    \begin{center}
        \includegraphics[width=.9\linewidth]{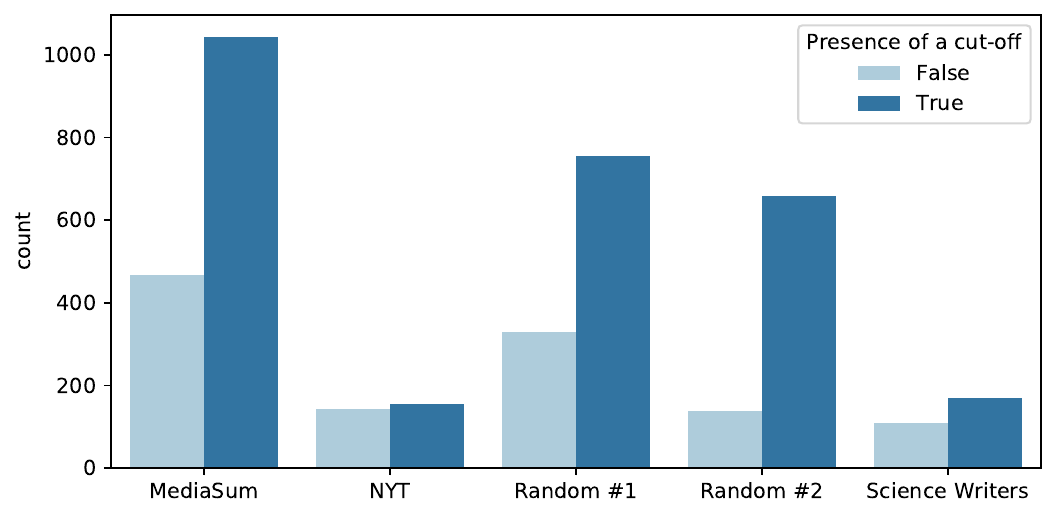}\vspace{-5pt}
        \caption{Amount of egos with and without a cut-off point.}
        \label{fig:ratio-cutoff} \vspace{-20pt}
    \end{center}
\end{figure}

To summarize, the algorithm returns a value called $\hat{n}$ that corresponds to $n_a$ if there is a cut-off point, and to $n_f$ if there is not.
With this algorithm, we can separate the egos into two groups: those that have a mature ego network (i.e., those for which we have been able to extract a cut-off in the saturation curve) and those that do not. The number of egos in the first and second groups is shown in Figure~\ref{fig:ratio-cutoff} for our datasets. It appears that in all datasets, and especially in the largest ones (MediaSum and both random datasets), egos with mature ego networks are the vast majority. In the rest of our analysis, we will retain only them, so that we can study their structural properties. 

\begin{algorithm}[t]
\caption{Find the cut-off point of the saturation curve}\label{alg:cut-off}

 \hspace*{\algorithmicindent} \textbf{Input: } $\mathbf{t} = \{i : t_i \in \mathcal{T}^n\}$ and $\mathbf{w} = \{s(t_i) : t_i \in \mathcal{T}^n\}$, i.e. the datapoints of the saturation curve \\
  \hspace*{\algorithmicindent} \textbf{Output: } $\hat{n}$, i.e. the cut-off point.\\
\begin{algorithmic}[1]
\State $\hat{n} \gets \Call{RecursiveCutOff}{\mathbf{t}, \mathbf{w}}$
\item[]
\Function{RecursiveCutOff}{$\mathbf{x}, \mathbf{y}$}
\If{\Call{IsLinear}{$\mathbf{x}, \mathbf{y}$}}
    \State \Return last element of $\mathbf{x}$
\Else
    \State $\hat{\mathbf{x}}, \hat{\mathbf{y}} \gets $ \Call{SplitSaturationCurve}{$\mathbf{x}, \mathbf{y}$}
    \State \Return \Call{RecursiveCutOff}{$\hat{\mathbf{x}}$, $\hat{\mathbf{y}}$}
 \EndIf
\EndFunction
\item[]
\Function{SplitSaturationCurve}{$\mathbf{x}, \mathbf{y}$}
    \Statex \hspace{2em}  \Comment{Subsetting notation ``$[:\!n]$'' means from first to $n$-th element}
        \Statex \hspace{2em}  \Comment{``$[n\!:]$'' means from $n$-th element to last}
    \State \Var{best_n} $\gets 1$
    \State \Var{lowest_rss} $\gets +\infty$
    \For{$n=1$ \textbf{to} $\max(y)-1$}
        \Statex \hspace{2em} \Comment{get RSS from standard least-squares regression}
        \State \Var{rss$_1$} $\gets$ \Call{LinearFit}{$\mathbf{x}[:\!n]$, $\mathbf{y}[:\!n]$} 
        \State \Var{rss$_2$} $\gets$ \Call{LinearFit}{$\mathbf{x}[n+1\!:]$, $\mathbf{y}[n+1\!:]$}
        \If{\Var{rss$_1$}+\Var{rss$_2$}$<$\Var{lowest_rss}}
            \State \Var{lowest_rss} $\gets$ \Var{rss$_1$}+\Var{rss$_2$}
            \State \Var{best_n} $\gets$ $n$
        \EndIf
    \EndFor
    \State \Return $\mathbf{x}[:\!best\_n]$,$\mathbf{y}[:\!best\_n]$
\EndFunction
\item[] 
\Function{IsLinear}{$\mathbf{x}, \mathbf{y}$}
    \Statex \hspace{2em} \Comment{$\beta_i$ is the Lasso coefficient associated with the polynomial term of degree $i$}
    \State $\beta_1, \ldots, \beta_p \gets $ \Call{LassoMaxVariableReduction}{$\mathbf{x}, \mathbf{y}$} 
    \If{$\beta_1 \neq 0$}
    \State \Return True
    \Else
    \State \Return False
    \EndIf
\EndFunction 
\end{algorithmic} 
\end{algorithm}

\vspace{-15pt}
\section{Results}
\label{sec:part2}


The goal of this section is to fully validate the methodology proposed in Section~\ref{sec:part1}. First, in Section~\ref{sec:optimal-circles-active} we show that the layered structure that was not present when considering the full ego network (Figure~\ref{fig:nb-circles-full}) emerges again when focusing on the active ego network, and we revisit its properties in Section~\ref{sec:revisiting-active-egonets}. Then we evaluate the robustness of the methodology to a varying amount of input data (Section~\ref{sec:robustness}). Finally, we show that active ego networks are stable over time (Section~\ref{sec:temporal-stability}). 

\vspace{-15pt}
\subsection{Optimal circle size for the active ego network}
\label{sec:optimal-circles-active}
\vspace{-5pt}

\begin{figure}[!t]
    \begin{center}
        \includegraphics[scale=.5]{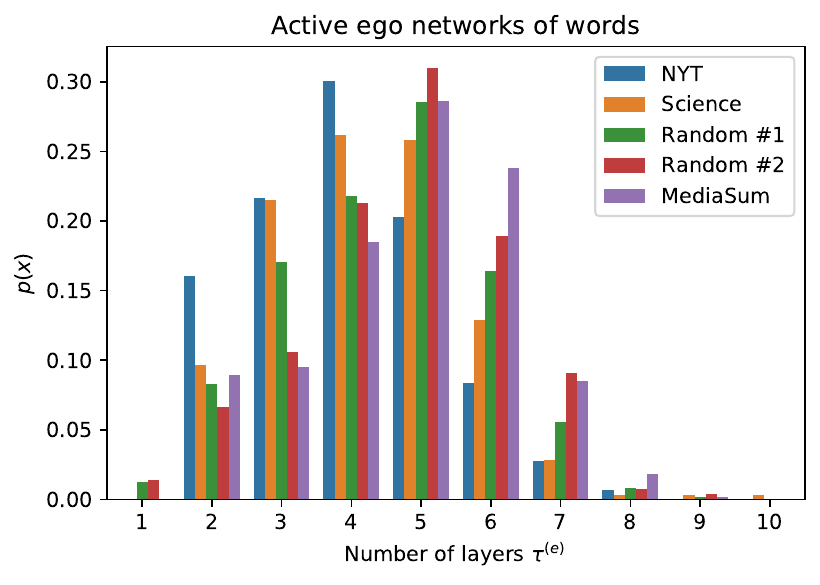}\vspace{-15pt}
        \caption{Distribution of the number of circles.}
        \label{fig:nb-circles-active} \vspace{-15pt}
    \end{center}
\end{figure}

We return to the initial motivation behind this work, namely the disappearance of the layered structure in the ego network of words within large textual corpora when failing to accurately identify the active portion of the ego network. This phenomenon was illustrated in Figure~\ref{fig:nb-circles-full}. By employing the methodology outlined in Section~\ref{sec:part1}, we can now effectively isolate\footnote{It is important to note, as mentioned earlier, that we exclude all egos that have not yet reached their saturation point to ensure that the observed ego networks are mature and not partially empty.} the active component of the ego network and ascertain whether the layered structure reemerges. Figure~\ref{fig:nb-circles-active} demonstrates that this is indeed the case. 
Comparing it with Figure~\ref{fig:nb-circles-full}, where the circles were computed on the full ego network, we observe that limiting the size of the ego network to the maximum number of actively used words shifts the mode from two circles to four or five circles, for all datasets. This means that the structure of the ego network fully emerges when the active part is properly isolated, similar to what happens for social ego networks. And that the methodology from Section~\ref{sec:part1} is able to properly identify the active part.

We now take a step further to demonstrate the effectiveness of the methodology proposed in Section~\ref{sec:active-egonet}. Given its recursive nature -- where the saturation curve undergoes successive cuts until identifying the linear phase and its corresponding cut-off point -- verifying the final cut's ability to produce a well-structured ego network is insufficient. Intermediate cut-off points, preceding the final recursive iteration, might still yield similarly well-structured ego networks. This discrepancy could imply imprecision in identifying the active segment of the ego network. To assess this, we analyse the evolving structure of ego networks at each iteration step using Algorithm~\ref{alg:cut-off}.
In Table~\ref{tab:iterations-recursive-method-mediasum}, we present the results for the Mediasum dataset exclusively, though readers interested in the results for other datasets can refer to the Supplemental Material (note that the observed trend is consistent across all datasets). Each row in Table~\ref{tab:iterations-recursive-method-mediasum} corresponds to egos with the same number of total iterations (one iteration for the first row, two for the second row, and so on). The emergence of a structured ego network is indicated by the distribution of the optimal number of circles, shifting its mode away from the value 2 (which signals a substantial lack of structure) as the final iteration is reached. Table~\ref{tab:iterations-recursive-method-mediasum} shows that a coherent structure only emerges at the last iteration of the proposed recursive method.
So, when we consider the results from Figure~\ref{fig:nb-circles-active} in conjunction with Table~\ref{tab:iterations-recursive-method-mediasum}, we not only demonstrate that our proposed method automatically leads to well-structured ego networks by excluding ``inactive'' words but also establish that such well-structured ego networks only emerge at the conclusion of the recursive steps.

\begin{table}[t]
    \centering
        \caption{Distribution of the optimal number of layers at each iteration of our recursive method on the Mediasum dataset. Each row contains egos with different numbers of total iterations, respectively 1, 2, and 3.}
    \label{tab:iterations-recursive-method-mediasum}
    \begin{tabular}{@{}cccc@{}} 
    \toprule
    \textbf{No cut} & \textbf{1st iteration} & \textbf{2nd iteration} & \textbf{3rd iteration} \\
    \midrule
        \includegraphics[width=0.22\linewidth]{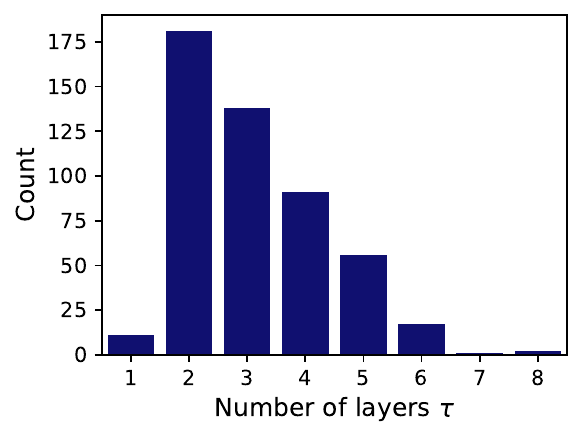} &  \includegraphics[width=0.22\linewidth]{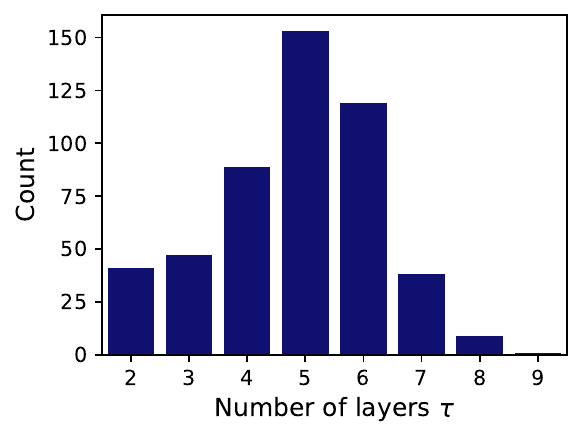} & & \\ \hline
         \includegraphics[width=0.22\linewidth]{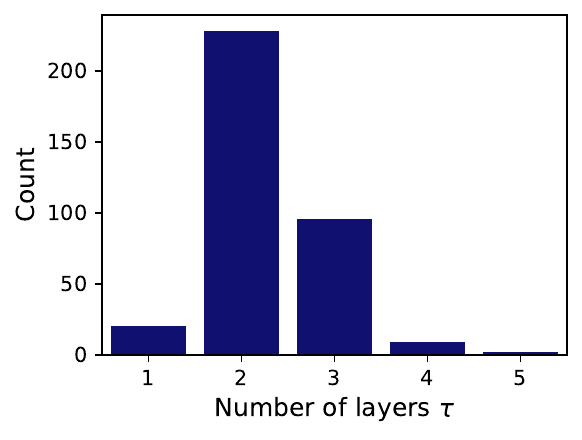} &  \includegraphics[width=0.22\linewidth]{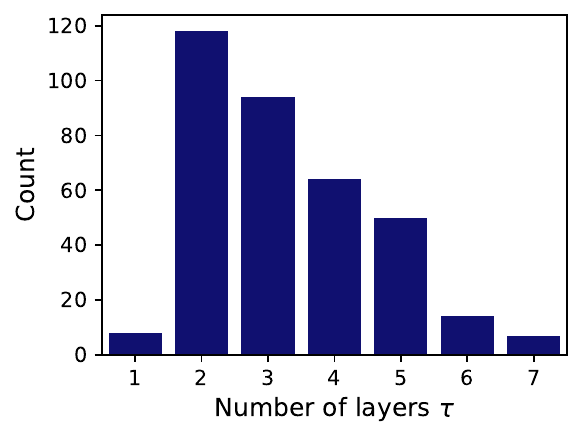} & 
         \includegraphics[width=0.22\linewidth]{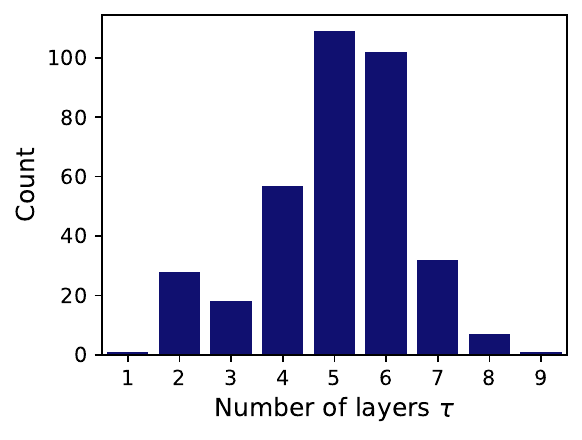} & \\ \hline
         \includegraphics[width=0.22\linewidth]{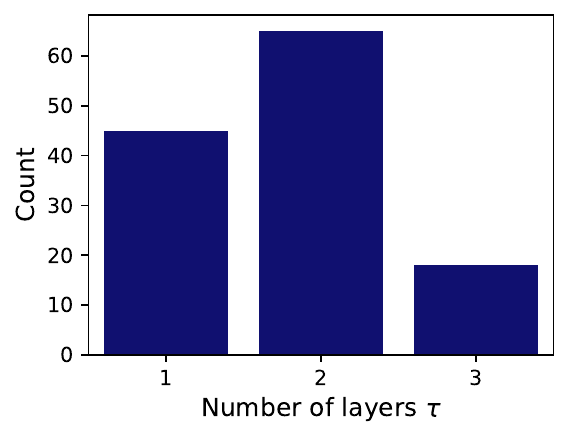} &  \includegraphics[width=0.22\linewidth]{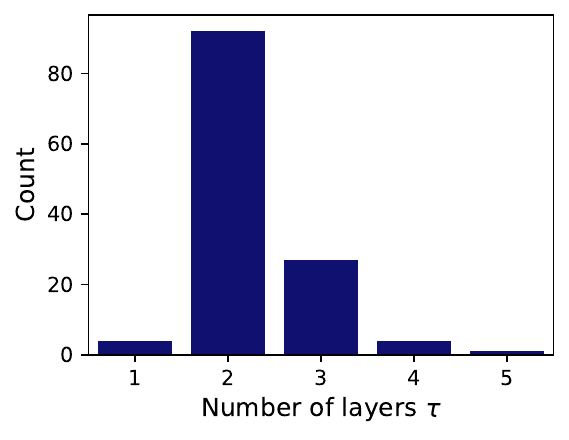} & 
         \includegraphics[width=0.22\linewidth]{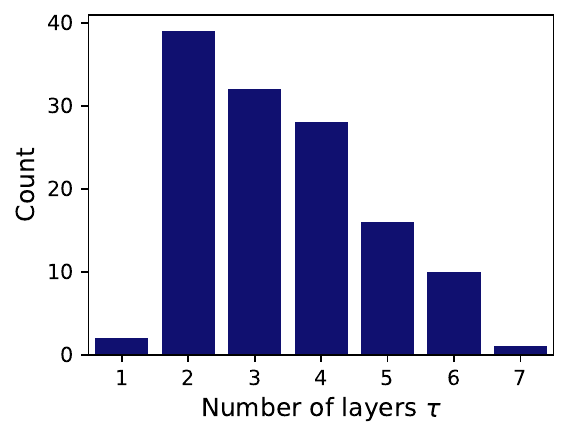} & 
         \includegraphics[width=0.22\linewidth]{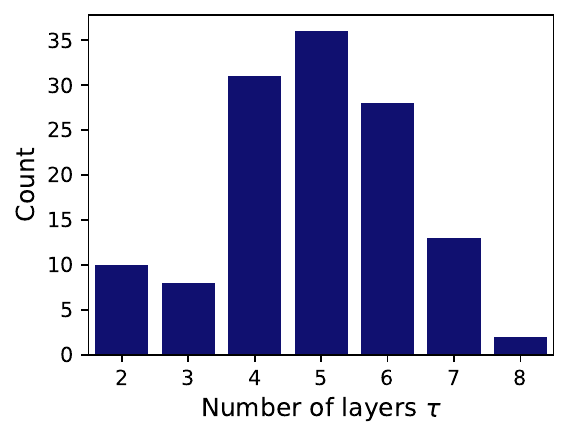}\\ \bottomrule
    \end{tabular} \vspace{-15pt}
\end{table}

\subsection{Revisiting the structural properties of word ego networks}
\label{sec:revisiting-active-egonets}

We can now investigate the properties of the active ego networks of words for the users in the datasets discussed in Section~\ref{sec:dataset}. Recall that egos that have not reached their cut-off point are excluded from the following analysis. The remaining ego networks are reduced to their active size $w^{n_a}$ obtained with the method of Section~\ref{sec:methodology}. From now on, we simplify the notation $w^{n_a}$ to $w$.

The analysis in Figure~\ref{fig:nb-circles-active} revealed that active ego networks typically consist of between 4 and 5 circles. It is worth noting that NYT journalists and science writers tend to have slightly fewer circles compared to random users and speakers in the MediaSum dataset. Notably, the ego networks of MediaSum speakers closely align with those of generic Twitter users \#2. Interestingly, a similar optimal range of 4 to 5 circles was also observed in the social domain~\cite{dunbar2015structure}. 

\begin{figure}
     \centering
     \subfloat[Plain]{
         \includegraphics[width=.9\linewidth]{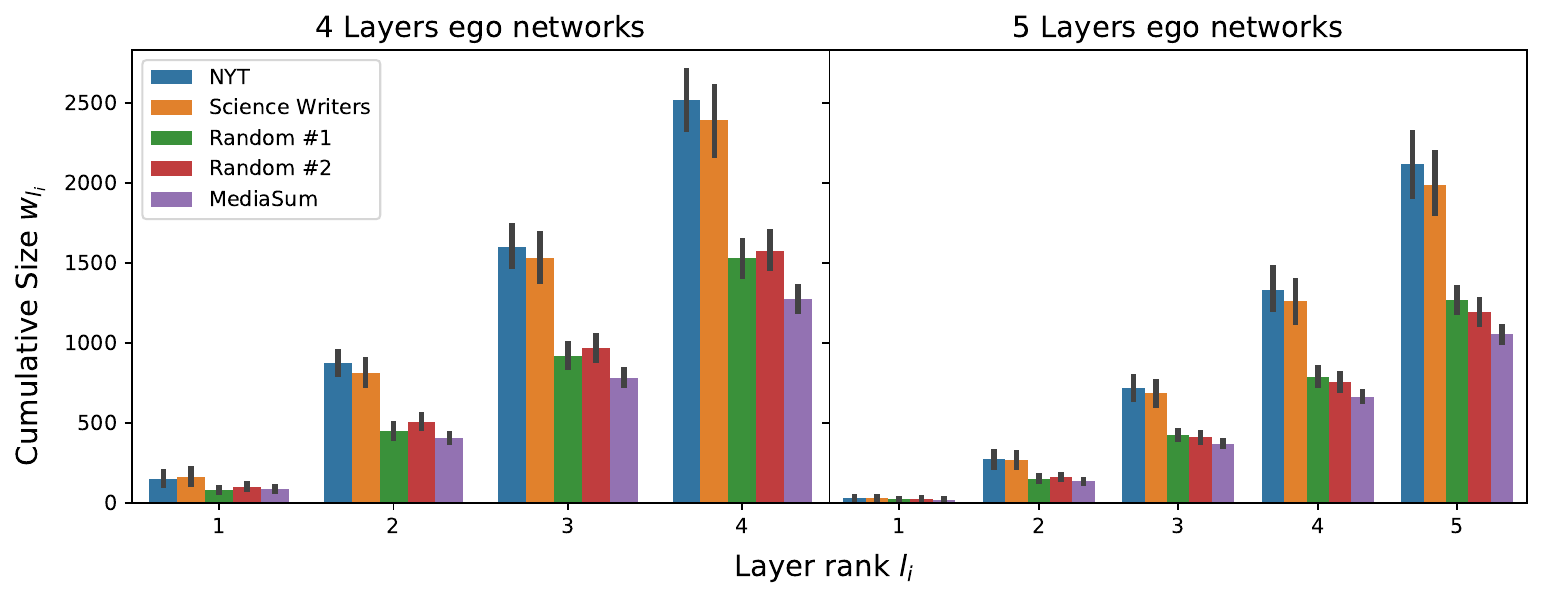}
         \label{fig:cumul-circle-size}}
     \hfill
     \subfloat[Normalized]{
         \includegraphics[width=.9\linewidth]{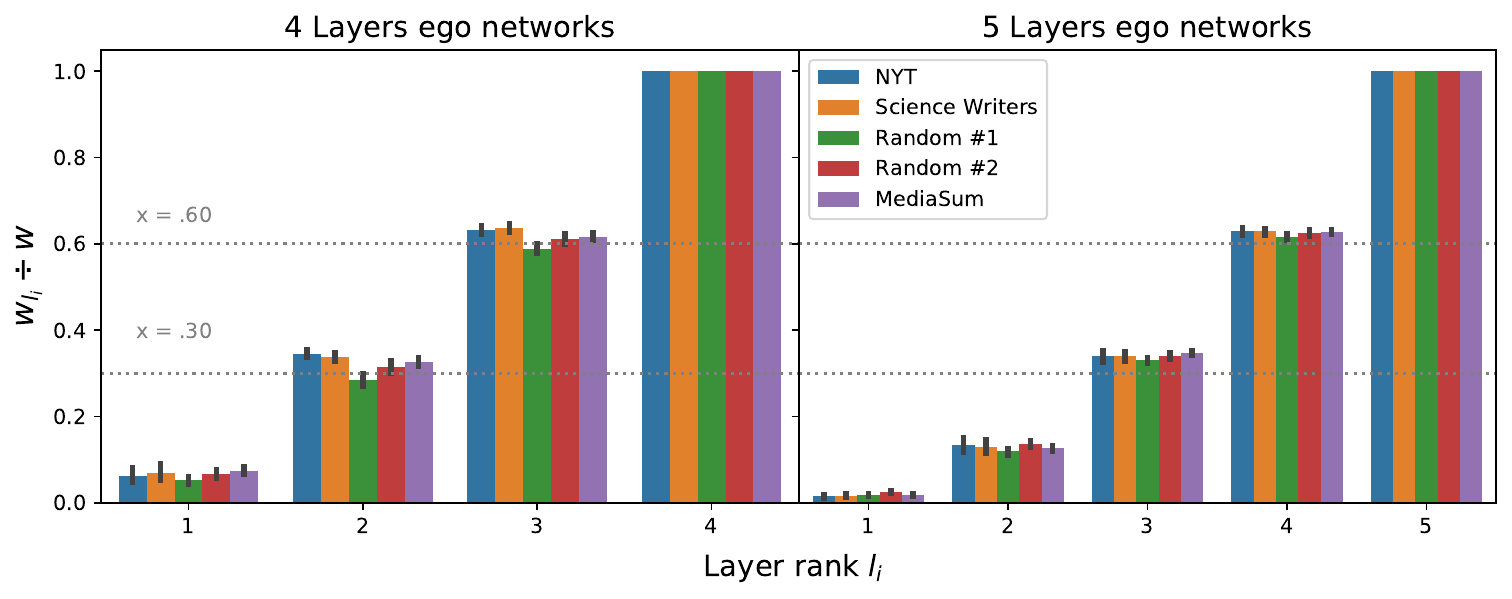}
         \label{fig:cumul-circle-normalized}
     }
        \caption{Layer size}
        \label{fig:three graphs}
        \vspace{-10pt}
\end{figure}

We now focus on the size of the ego network layers. For this analysis, we consider four- and five-layered ego networks, which are the most frequent cases in the five datasets, as shown in Figure~\ref{fig:nb-circles-active}, hence providing more samples for statistical reliability. In Figure~\ref{fig:cumul-circle-size}, the average layer sizes $w_{l_i}$ are ranked from the innermost ($l_1$) to the outermost one ($l_4$ or $l_5)$. Recall that the active size of an ego network, which corresponds to the total number of unique words before the cut-off, is also the size of the outermost layer. The layers of the ego networks from specialized Twitter datasets (NYT journalists and science writers) are, on average, bigger compared to random users and MediaSum speakers. Again, MediaSum speakers are quite well aligned with generic users on Twitter.
According to the saturation curve methodology in Section~\ref{sec:active-egonet}, it means that they can handle a larger number of words before saturating their ability to bring new ones into their active vocabulary. The size of five-layered ego networks is consistently lower compared to the four-layered ones ($\sim$20\% lower independently of the dataset). However, it seems that words have a similar distribution across the layers regardless of the dataset. We verify this property in the following.

We define the normalized layer size as the ratio between the layer size and the ego network size $\frac{w_{l_i}}{w}$. As can be seen in Figure~\ref{fig:cumul-circle-normalized}, normalized layer sizes are very similar across datasets. The penultimate layer $l_{\tau-1}$ consistently accounts for 60\% of the ego network size, and the second to last layer $l_{\tau-2}$ accounts for 30\%:
\begin{equation}
    \begin{cases}
        \cfrac{w_{l_{\tau-1}}}{w} \simeq 0.6\\
        \cfrac{w_{l_{\tau-2}}}{w} \simeq 0.3
    \end{cases}
    \label{eq:cumulative-ratio-invariant}
\end{equation}
We can observe the same pattern in the case of six-layered ego networks as well as for the penultimate layer of three-layered ego networks (Table~\ref{table:coeff}).
These values are very similar to those obtained in our previous paper \cite{ollivier2020}, where the average ego network size was smaller. This means that the main difference between two ego networks with different numbers of layers is in the organisation of the inner layers. Note also that this regularity applies to all datasets, with no remarkable difference, further supporting the cross-domain generalizability of the ego network of words model.


\begin{table}[!t]
\center
\caption{Average ratio between a layer size $w_{l_i}$ and the active size of the ego network $w$ , in all datasets.
\label{table:coeff}}
\scriptsize
\begin{tabular}{l l c c c c c c}
\toprule
\multirow{2}{*}{\textbf{Dataset}} & \multirow{2}{*}{\textbf{\# of layers}} & \multicolumn{6}{c}{\textbf{Layer Rank}}           \\ \cmidrule(r){3-8} 
                                  &  & \textbf{1}    & \textbf{2}    & \textbf{3}    & \textbf{4}    & \textbf{5}    & \textbf{6}   \\ \midrule

\multirow{4}{*}{NYT} & 3 layers & .16 & \cellcolor{gray!50}.55 & 1 &  &  &  \\
& 4 layers & .05 & \cellcolor{gray!25}.32 & \cellcolor{gray!50}.61 & 1 &  &  \\
& 5 layers & .01 & .11 & \cellcolor{gray!25}.33 & \cellcolor{gray!50}.62 & 1 &  \\
& 6 layers & .00 & .03 & .15 & \cellcolor{gray!25}.34 & \cellcolor{gray!50}.63 & 1 \\
\midrule
\multirow{4}{*}{Science Writers} & 3 layers & \cellcolor{gray!25}.25 & \cellcolor{gray!50}.58 & 1 &  &  &  \\
& 4 layers & .06 & \cellcolor{gray!25}.33 & \cellcolor{gray!50}.62 & 1 &  &  \\
& 5 layers & .01 & .14 & \cellcolor{gray!25}.33 & \cellcolor{gray!50}.63 & 1 &  \\
& 6 layers & .01 & .03 & .15 & \cellcolor{gray!25}.34 & \cellcolor{gray!50}.63 & 1 \\
\midrule
\multirow{4}{*}{Random \#1} & 3 layers & .11 & \cellcolor{gray!50}.53 & 1 &  &  &  \\
& 4 layers & .04 & \cellcolor{gray!25}.24 & \cellcolor{gray!50}.58 & 1 &  &  \\
& 5 layers & .01 & .10 & \cellcolor{gray!25}.30 & \cellcolor{gray!50}.60 & 1 &  \\
& 6 layers & .00 & .03 & .14 & \cellcolor{gray!25}.33 & \cellcolor{gray!50}.62 & 1 \\
\midrule
\multirow{4}{*}{Random \#2} & 3 layers & .13 & \cellcolor{gray!50}.55 & 1 &  &  &  \\
& 4 layers & .05 & \cellcolor{gray!25}.26 & \cellcolor{gray!50}.59 & 1 &  &  \\
& 5 layers & .02 & .12 & \cellcolor{gray!25}.33 & \cellcolor{gray!50}.63 & 1 &  \\
& 6 layers & .00 & .03 & .12 & \cellcolor{gray!25}.33 & \cellcolor{gray!50}.61 & 1 \\
\midrule
\multirow{4}{*}{MediaSum} & 3 layers & .15 & \cellcolor{gray!50}.56 & 1 &  &  &  \\
& 4 layers & .06 & \cellcolor{gray!25}.31 & \cellcolor{gray!50}.61 & 1 &  &  \\
& 5 layers & .01 & .11 & \cellcolor{gray!25}.34 & \cellcolor{gray!50}.63 & 1 &  \\
& 6 layers & .00 & .03 & .14 & \cellcolor{gray!25}.34 & \cellcolor{gray!50}.63 & 1 \\
\bottomrule
\end{tabular} \vspace{-15pt}
\end{table}

\begin{figure}[!t]
    \begin{center}
        \includegraphics[width=\linewidth]{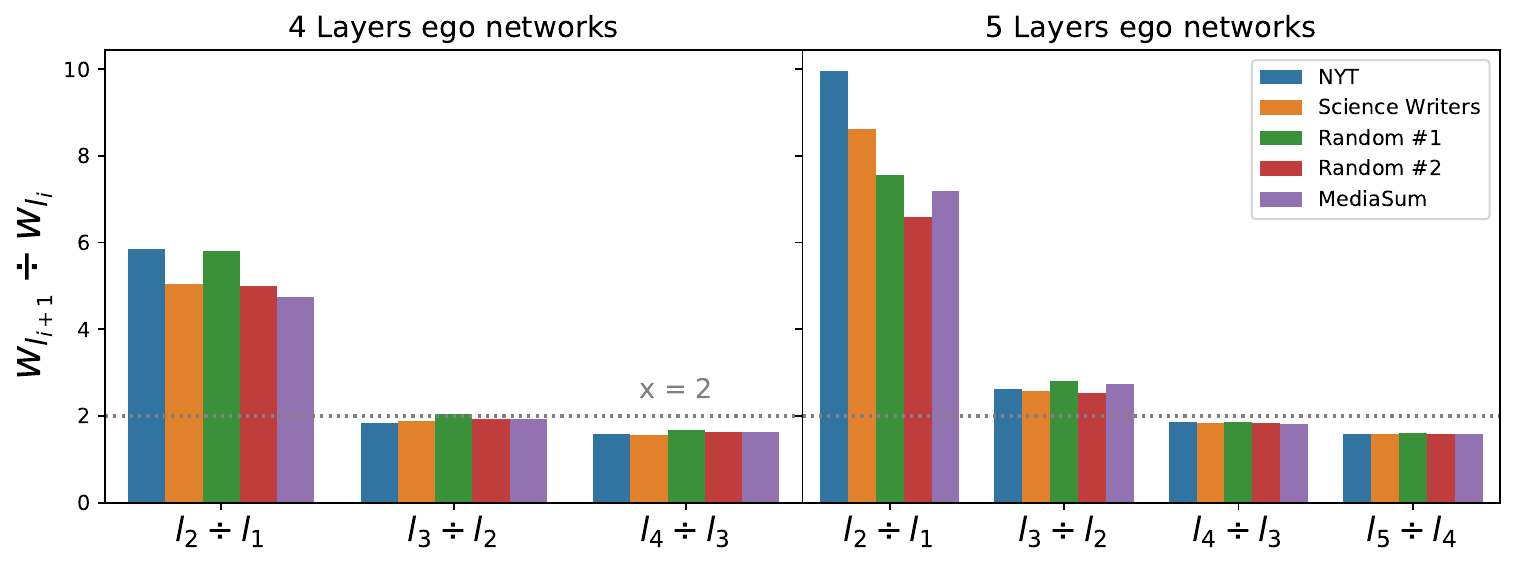}\vspace{-15pt}
        \caption{Scaling ratio.}
        \label{fig:cumul-circle-scaling-ratio}
    \end{center}
    \vspace{-25pt}
\end{figure}

The scaling ratio is a metric that describes how the layer size grows from a layer $l_{i-1}$ to the outer layer $l_i$: $\cfrac{w_{l_i}}{w_{l_{i-1}}}$.
As we can see in Figure~\ref{fig:cumul-circle-scaling-ratio} the ratio is very similar across the datasets for $i \geq 3$. The ratio tends to reach a value slightly below two toward the outermost layers. These results are the same as those obtained in the paper \cite{ollivier2020}. 

When comparing the current findings with previous research~\cite{ollivier2020,ollivier2022} that focused on ego networks of words, we must consider two aspects: first, the current work is based on more diverse and larger datasets, and second, the previous work did not specifically focus on the active network segment of the ego network (because a robust methodology for identifying it did not exist). Despite these considerations, the observations in the previous work~\cite{ollivier2020,ollivier2022} surprisingly align well with the current findings, particularly concerning the number of circles (which were found to be between 5 and 7 in~\cite{ollivier2022} vs 4-5 in this work) and the scaling ratio (approximately the same in~\cite{ollivier2022}). However, when examining the absolute sizes of individual layers, we notice larger sizes in this work compared to~\cite{ollivier2022}. To better understand this behavior, we can focus on the Twitter datasets, which are common to both studies (same users, shorter timelines in~\cite{ollivier2022}). Both the similarities and differences in the ego networks can be explained by the fact that the observed timelines in~\cite{ollivier2022} generally cover around or slightly less than the cut-off point. Consequently, the ego network structure becomes apparent, but some words are missing to make it fully complete (hence the smaller layers). Vice versa, the timelines we use in the current study cover much more than the cut-off point, hence, without a proper methodology to identify the active network, the resulting structure is meaningless (as shown in Section~\ref{sec:full-egonet}). Note that the slightly higher number of optimal circles in~\cite{ollivier2022} can similarly be explained by an observation window below the cut-off point. While this may appear counterintuitive, the number of circles tends to grow as the number of data points decreases. This occurs because the clustering algorithm may detect spurious groupings when data points become more scattered.

\vspace{-15pt}
\subsection{Robustness of the methodology}
\label{sec:robustness}

In this section and the subsequent one, our primary focus lies on internally validating the proposed methodology for identifying the active network. We start with an analysis of the robustness of the methodology to the amount of available data. Specifically, the cut-off point of the active ego network should be a characteristic of each ego and not dependent on the size of the ego data fed to the algorithm. This implies that our algorithm should consistently determine the same cut-off point for a given ego, except when there is insufficient data to reach that point. In this section, we verify that this is the case. 

Let us consider a tagged ego $e$ whose saturation curve contains a cut-off point $n_a$. Recall that $\mathcal{T}^n \subseteq \mathcal{T}$ and $\mathcal{W}^n \subseteq \mathcal{W}$, for any $n < n^f$. 
When \textsc{RecursiveCutOff} in Algorithm~\ref{alg:cut-off} is fed $\mathcal{T}^{n}$ and $\mathcal{W}^{n}$ where $n < n^f$, it should return $n_a$ if $n \geq n_a$ and $n$ otherwise (if $n$ is below the cut-off there is no cut-off to find). As $n$ grows, then, the corresponding size of the active ego network will grow. When $n$ reaches $n_a$, the active ego network is mature and should not grow anymore.
%
%
%
%
This means that the active network size $\hat{w}^{n}$ for varying $n$ should follow the ideal behavior :
\begin{equation}
\hat{w}^{n} = 
    \begin{cases}
        w^{n} \;\;\;\text{when} \;\;\; n \in [0,n_a]\\
        w^{n_a} \;\;\;\text{when}\;\;\; n \in [n_a,n_f]
    \end{cases}
    \label{eq:robustness-eq2}
\end{equation}
%

\begin{figure*}[t]
    \begin{center}
        \includegraphics[width=\linewidth]{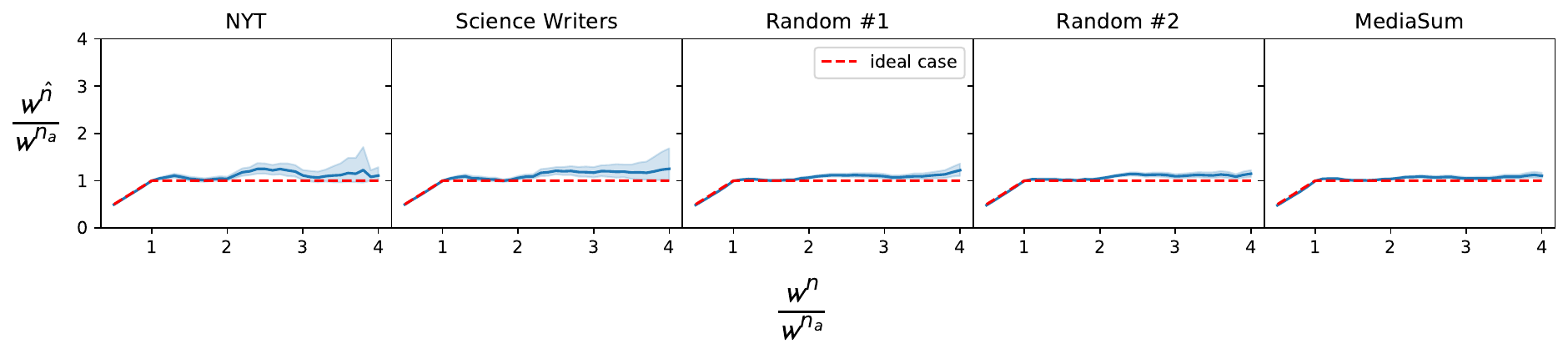}\vspace{-25pt}
        \caption{The stability of the algorithm is close to the ideal case.}
        \label{fig:stability}
        \vspace{-10pt}
    \end{center}
\end{figure*}

        
%
In Fig~\ref{fig:stability} we plot the ratio $\cfrac{\hat{w}^{n}}{w^{n_a}}$. We expect $\cfrac{\hat{w}^{n}}{w^{n_a}}$ to grow from zero to one and then remain stable around one (implying that for any $n > n_a$, the calculated cut-off remains the same, regardless of the increasing size of the data being fed to the algorithm). Fig~\ref{fig:stability} confirms that the behavior of the calculated cut-off, and hence of the resulting size of the active network, is close to the ideal case in every dataset, despite some noise due to a lower number of ego networks in the NYT journalists and science writers datasets.

\vspace{-10pt}
\subsection{Temporal stability of the active network size}
\label{sec:temporal-stability}

With the methodology introduced in Section~\ref{sec:part1}, we are able to extract the active size of an ego network of words with respect to an observed tuple of tokens $\mathcal{T}$. This size corresponds to the volume of words actively used by the ego and whose boundary is associated with token $t_{n_a}$  (from which the use of new words becomes rare).  However, this count assumes that a word used at the beginning of $\mathcal{T}$ is still part of the active ego network. This raises the question of what would happen if we had started observing the language production of a speaker/user not from token $t_0$ but from a generic token $t_{\delta}$. 
By shifting the start of the analysis from $t_0$ to $t_{\delta}$, we study
the dynamic evolution of the size of the active network, which is important because it allows us to assess whether the cognitive ability to add words to one's active vocabulary evolves over time. 

\begin{figure}[!t]
    \begin{center}
        \vspace{-10pt}
        \includegraphics[width=.6\linewidth]{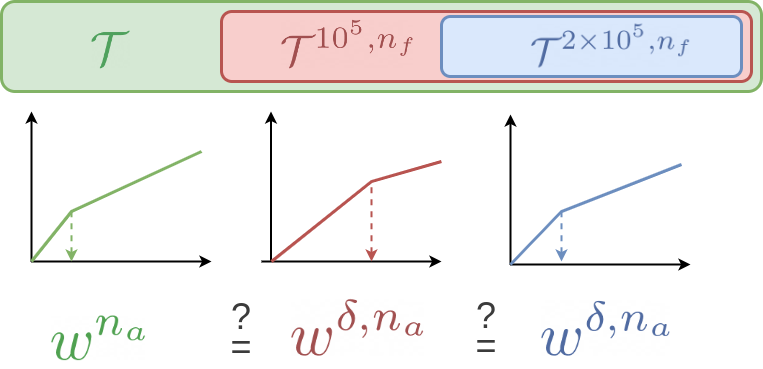}\vspace{-5pt}
        \caption{The diagram illustrates the temporal analysis procedure of the active size of an ego network. A temporal change corresponds here to a change in the index $\delta$ of the first word of the sequence used to build the ego network. This change leads by construction to a different saturation curve from which we will extract and study the variability of the active part size $w^{\delta,n_a}$.}
        \label{fig:temp-shift-stability-method}
    \end{center}
    \vspace{-15pt}
\end{figure}

\begin{figure}[!t]
    \begin{center}
        \includegraphics[width=.8\linewidth]{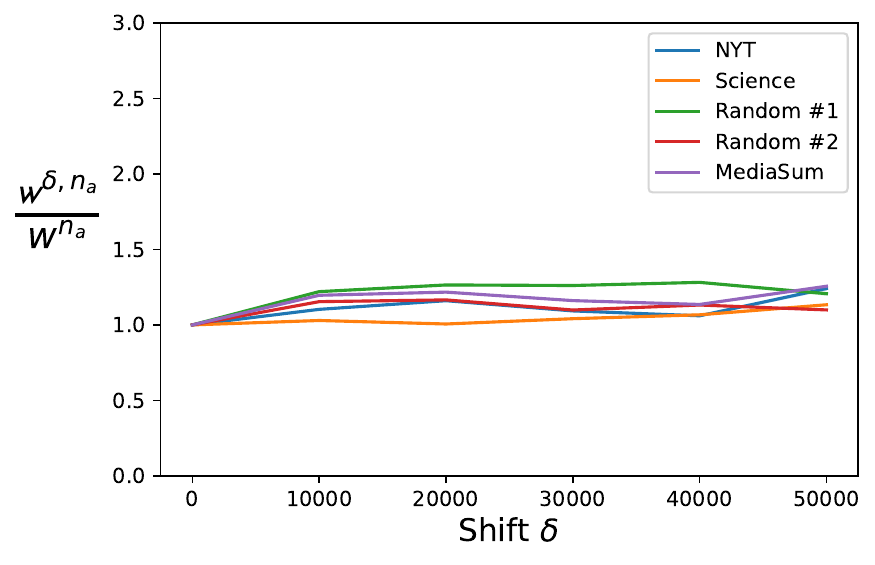} \vspace{-10pt}
        \caption{The shift $\delta$ of the token sequence from which the ego networks are built has almost no influence on the active size $w^{\delta,n_a}$ on. In order to average that behaviour at the dataset level, we consider the ratio $\cfrac{w^{\delta,n_a}}{w^{n_a}}$ where the divisor is the original active size  ($\delta=0$). This ratio is consistently close to one (the maximum average value is $1.25$, reached by the MediaSum dataset for $\delta=5 \times 10^5$). These aggregated values are reliable since the 95\% average confidence interval is only $\pm0.08$.}
        \label{fig:temp-shift-stability-res}
    \end{center}
    \vspace{-15pt}
\end{figure}

To evaluate the temporal evolution of the active network size, we change the starting index of the sequence of tokens $\mathcal{T}^{n_f}$ from which we build the saturation curve. We call that shift $\delta$, the updated tuple of tokens $\mathcal{T}^{\delta,n_f}$ and the corresponding word tuple $\mathcal{W}^{\delta,n_f}$.  We build a new saturation curve, from which we extract an active network size $w^{\delta,n_a}$ (Figure~\ref{fig:temp-shift-stability-method}). We want to compare  $w^{\delta,n_a}$, when $\delta$ varies, against the original active size $w^{n_a}$. If $w^{\delta,n_a}$ remains comparable to the second, it means that the active size of the network is stable over time. 
Thus, in the following, we study the ratio $\cfrac{w^{\delta,n_a}}{w^{n_a}}$.
Note that the more we shift $\delta$ the more we run the risk of not observing egos for enough time and, consequently, of not having mature ego networks (much like the situation in which no cut-off could be found in Section~\ref{sec:active-egonet}). Thus, when shifting with $\delta$ we always make sure that, for each ego, at least $n_a$ tokens are observed. This means that we operate in the range $\delta \in [0, \delta_{max}]$, with $\delta_{max}=n_f - n_a$. Note also that, differently from the previous section, here we never operate below the cut-off point $n_a$.
%
In Figure~\ref{fig:temp-shift-stability-res}, we choose a $\delta$ range from 0 to $5 \cdot 10^4$. That maximum was chosen because it is the largest value for which at least 25\% of the ego network has a $\delta_{max}$ higher than it. 

Following the above methodology, in Figure~\ref{fig:temp-shift-stability-res} we plot $\cfrac{w^{\delta,n_a}}{w^{n_a}}$ as a function of $\delta$.
We can observe that the ratio (hence, the size of the active ego network) remains stable when $\delta$ grows, independently of the dataset. This supports our hypothesis that the size and internal structure of the ego network are bound by cognitive constraints that are applied at different intensities depending on the individual, but which are themselves stable over time. 

\section{Conclusion}
\label{sec:conclusion}


In this work, we investigated the cognitive limitations in human language production and presented the ego network of words as a model to capture structural properties associated with these constraints. The paper introduces the concept of an ``active'' part of the ego network, which represents the words actively used by an individual, and demonstrates that beyond this active part, the structure of the ego network becomes poorly organized. A robust methodology is proposed to extract the active part of the ego network, and its effectiveness is validated using interview transcripts and tweets datasets. Restricting our analysis to the active part of the ego networks, as commonly done when analyzing ego networks in the social domain, we have confirmed that the structural properties of the ego network of words, such as the number of circles and the scaling ratio between circles, are consistent across different domains. 
In this paper, we have provided a method for classifying words into concentric circles when enough words are observed to fill the active ego network. As future work, we could explore the application of supervised classification techniques, such as neural networks. Specifically, these techniques could be used to predict which circle a word falls into when only a small sample of words is available for a given person.

The presented methodology and findings have implications for various fields, including linguistics, cognitive science, and social network analysis. For the latter, we could intersect our research with studies focusing on social dynamics. More specifically, if we build ego networks of words based solely on text produced during conversations between the ego and other persons, we could explore the link between the position of words used with a given interlocutor in the ego network of words and the interlocutor's position in the social ego network~\cite{dunbar2015structure,arnaboldi2012analysis}. Interpersonal language inherently contains social information about the level of intimacy~\cite{locher2010interpersonal}, as demonstrated by~\cite{pei2020quantifying}, whose authors leveraged levels of uncertainty and swearing to predict the familiarity between interlocutors. Thus, an ego network of words could serve as a microscope to identify structural and semantic features of language in specific contexts of use.
Moreover, it is established that conventional linguistic features (such as topic and sentiment) extracted from online social media offer valuable insights into the socio-economic status of the writer~\cite{preoctiuc2015studying}. Consequently, we could employ our model to investigate whether indicators of socioeconomic status are manifested differently across various ego network circles.

\bibliographystyle{ieeetr}  
\bibliography{bibliography}  


\vskip -2\baselineskip plus -1fil
\begin{IEEEbiography}[{\includegraphics[width=1in,height=1.25in,clip,keepaspectratio]{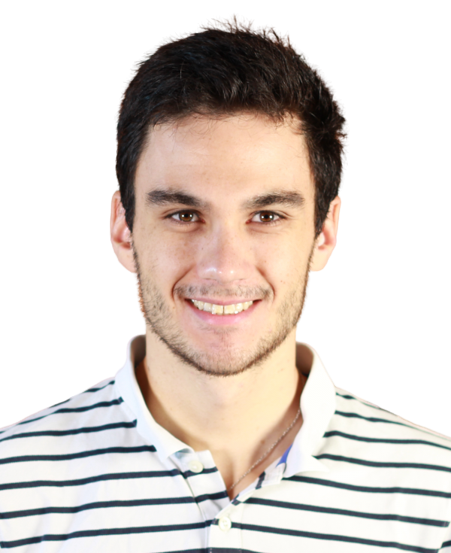}}]{Kilian Ollivier}
 completed his PhD in Data Science at IIT-CNR, and currently works as a data scientist at CBTW in Paris (France). He graduated in 2017 as a software engineer from INSA Lyon (France) and obtained an MSc in computer science from the University of Passau (French-German agreement for a double diploma). He started his research on data-driven discovery of human cognitive limits in early 2018 as a research fellow at IIT-CNR, then pursued his research as part of the PhD program.
\end{IEEEbiography}

\vskip -2.5\baselineskip plus -1fil

\begin{IEEEbiography}[{\includegraphics[width=1in,height=1.25in,clip,keepaspectratio]{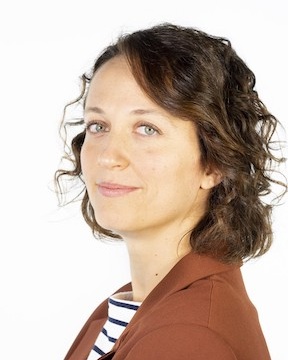}}]{Chiara Boldrini}
is a Senior Researcher at IIT-CNR. Her research interests are in decentralized AI, computational social sciences, mobile and ubiquitous systems. She has published 60+ papers on these topics. She is the IIT-CNR co-PI for H2020 SoBigData++ and HumaneE-AI-Net projects, and was involved in several EC projects since FP7. She currently holds the position of Editor-in-Chief for Special Issues at Elsevier Computer Communications and is a member of the Editorial Board of Elsevier Pervasive and Mobile Computing. She served as TPC chair of IEEE PerCom'24 and, over the years, has been in the organizing committee of several IEEE and ACM conferences/workshops, including IEEE PerCom and ACM MobiHoc.
\end{IEEEbiography}

\vskip -2.5\baselineskip plus -1fil

\begin{IEEEbiography}[{\includegraphics[width=1in,height=1.25in,clip,keepaspectratio]{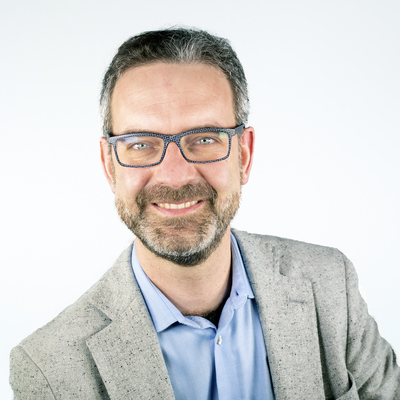}}]{Andrea Passarella} 
(PhD 2005) is a Research Director at the Institute of Informatics and Telematics (IIT) of the National Research Council of Italy (CNR). Prior to joining IIT, he was with the Computer Laboratory of the University of Cambridge, UK. He has published 200+ papers on Online and Mobile social networks, decentralised AI, Next Generation Internet, opportunistic, ad hoc and sensor networks, receiving the best paper award at IFIP Networking 2011 and IEEE WoWMoM 2013. He served as General Chair for IEEE PerCom 2022. He is the founding Associate Editor-in-Chief of Elsevier Online Social Networks. He is co-author of the book ``Online Social Networks: Human Cognitive Constraints in Facebook and Twitter Personal Graphs'' (Elsevier, 2015), and was Guest Co-Editor of several special sections in ACM and Elsevier Journals. He is the PI of the EU CHIST-ERA SAI (Social Explainable AI) project.
\end{IEEEbiography}

\vskip -2.5\baselineskip plus -1fil

\begin{IEEEbiography}[{\includegraphics[width=1in,height=1.25in,clip,keepaspectratio]{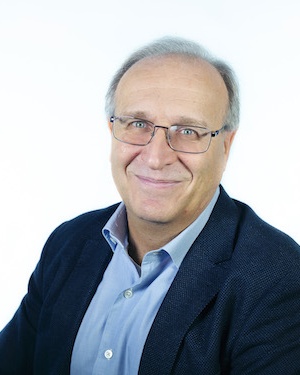}}]{Marco Conti}
is a CNR research director and, currently, he is the director of IIT-CNR institute. He has published more than 400 scientific articles related to design, modeling, and experimentation of computer and communication networks, pervasive systems, and online social networks. He is the founding EiC of Online Social Networks and Media, EiC for Special Issues of Pervasive and Mobile Computing, and, from 2009 to 2018, EiC of Computer Communications. He has received several awards, including the Best Paper Award at IFIP Networking 2011, IEEE ISCC 2012, and IEEE WoWMoM 2013. 
He was included in the "2017 Highly Cited Researchers" list compiled by Web of Science for the most cited articles in Computer Science. 
He served as the General/Program chair of several major conferences, including IFIP Networking 2002, IEEE WoWMoM 2005 and 2006, IEEE PerCom 2006 and 2010, ACM MobiHoc 2006, IEEE MASS 2007 and IEEE SmartComp 2021.
\end{IEEEbiography}









\end{document}


\appendix

\renewcommand\thetable{A\arabic{table}} 
\setcounter{table}{0}

\section*{Structure of ego networks at each iteration step}
\label{sec:appendix}

In this appendix, we provide the figures related to the emergence of the ego network structure at each iteration of the proposed recursive method for the datasets that could not be accommodated in the main paper. Table~\ref{tab:iterations-recursive-method-nyt} reports the results for the NYT dataset, Table~\ref{tab:iterations-recursive-method-science} for Science Writers, Table~\ref{tab:iterations-recursive-method-random1} for the Random \#1 dataset, and Table~\ref{tab:iterations-recursive-method-random2} for the Random \#2 dataset.

\begin{table}[h]
    \centering
        \caption{Distribution of the optimal number of layers at each iteration of our recursive method on the NYT dataset. Each row contains egos with different numbers of total iterations, respectively 0, 1, and 2.}
    \label{tab:iterations-recursive-method-nyt}
    \begin{tabular}{@{}ccc@{}} 
    \toprule
    \textbf{No cut} & \textbf{1st iteration} & \textbf{2nd iteration} \\
    \midrule
    \includegraphics[width=0.30\linewidth]{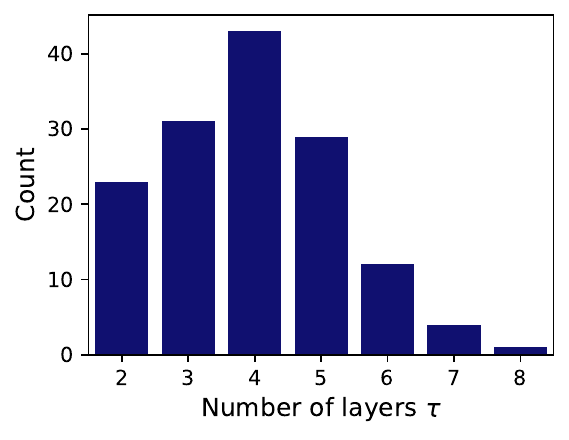} &   &\\ \hline
        \includegraphics[width=0.30\linewidth]{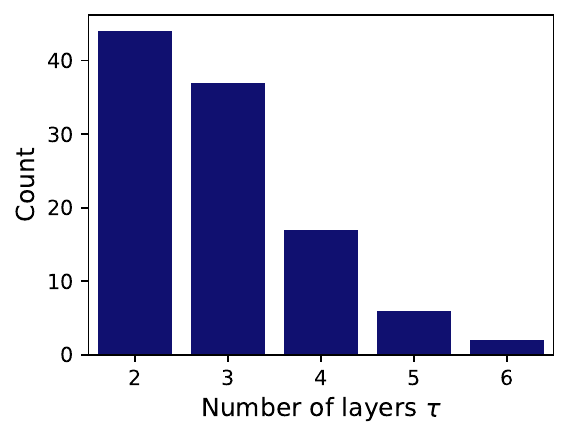} &  \includegraphics[width=0.30\linewidth]{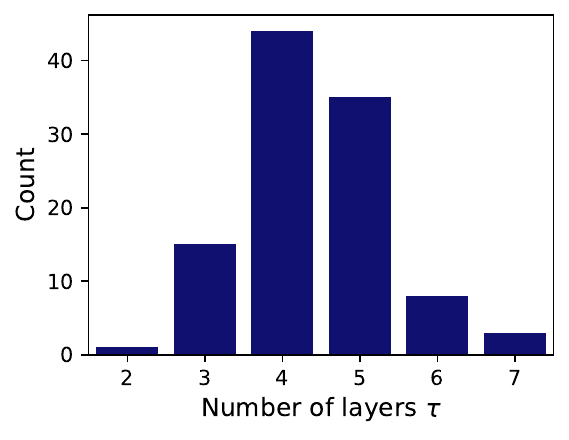} & \\ \hline
         \includegraphics[width=0.30\linewidth]{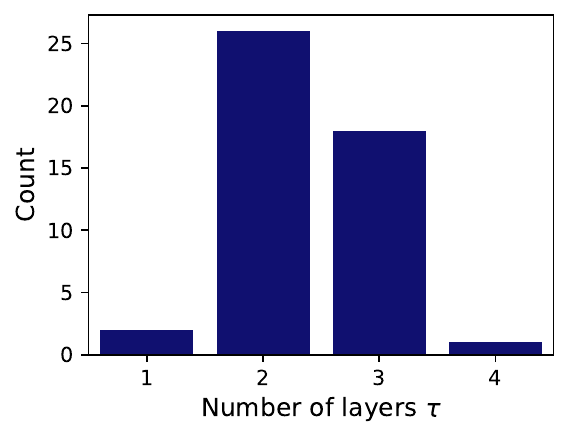} &  \includegraphics[width=0.30\linewidth]{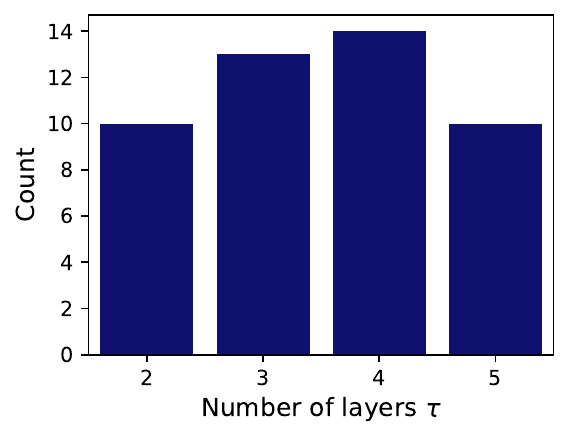} & 
         \includegraphics[width=0.30\linewidth]{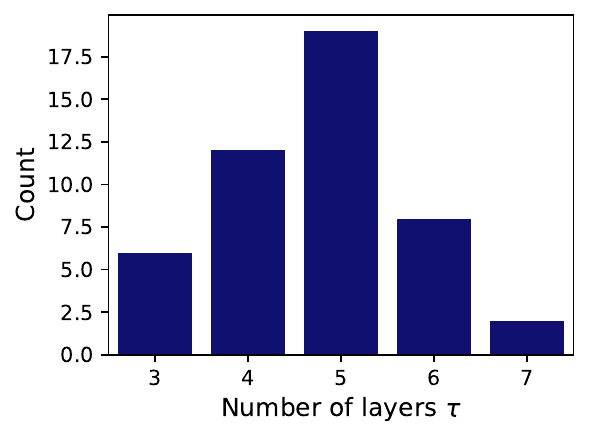} \\
         \bottomrule
    \end{tabular}
\end{table}

\begin{table}[h]
    \centering
    \caption{Distribution of the optimal number of layers at each iteration of our recursive method on the Science Writers dataset. Each row contains egos with different numbers of total iterations, respectively 0, 1, and 2.}
    \label{tab:iterations-recursive-method-science}
    \begin{tabular}{@{}ccc@{}} 
    \toprule
    \textbf{No cut} & \textbf{1st iteration} & \textbf{2nd iteration} \\
    \midrule
    \includegraphics[width=0.30\linewidth]{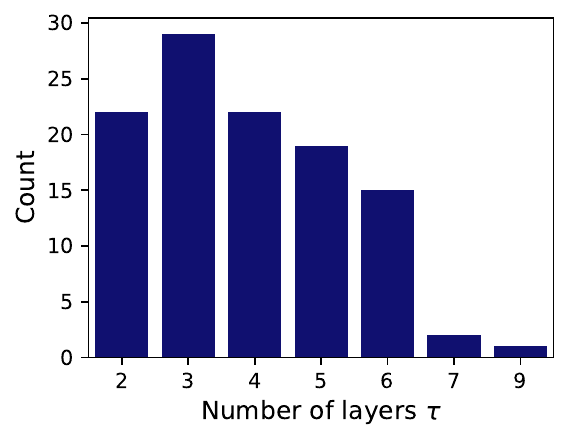} &   &\\ \hline
        \includegraphics[width=0.30\linewidth]{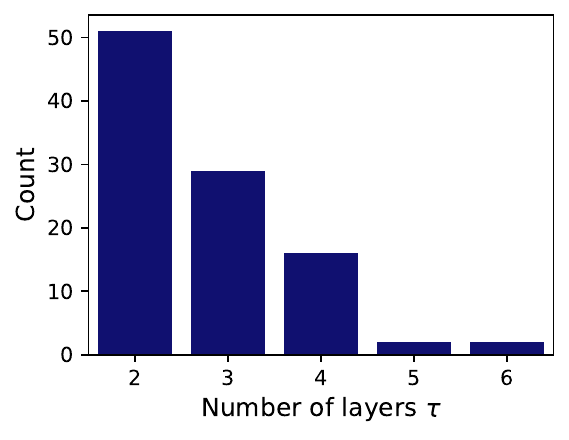} &  \includegraphics[width=0.30\linewidth]{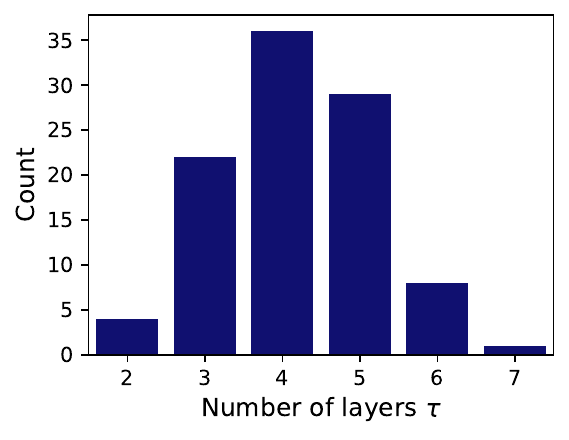} & \\ \hline
         \includegraphics[width=0.30\linewidth]{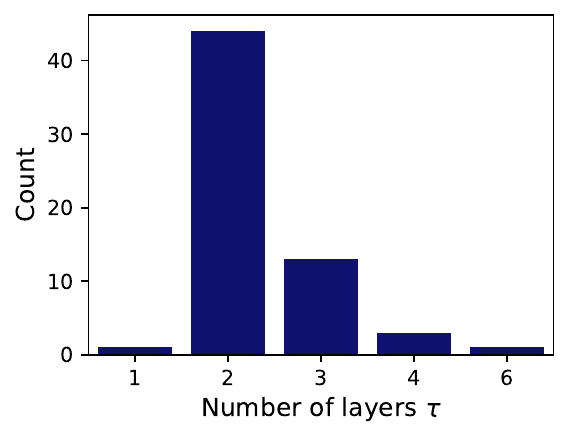} &  \includegraphics[width=0.30\linewidth]{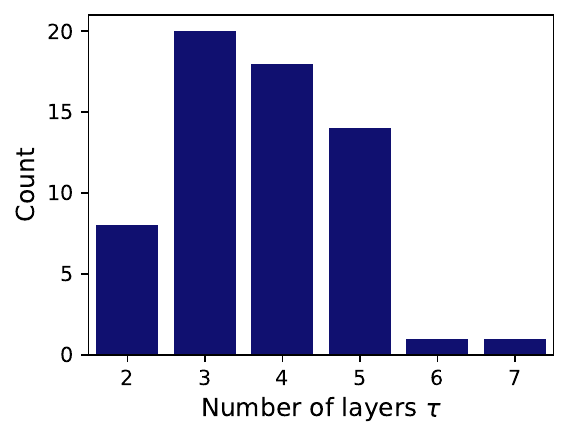} & 
         \includegraphics[width=0.30\linewidth]{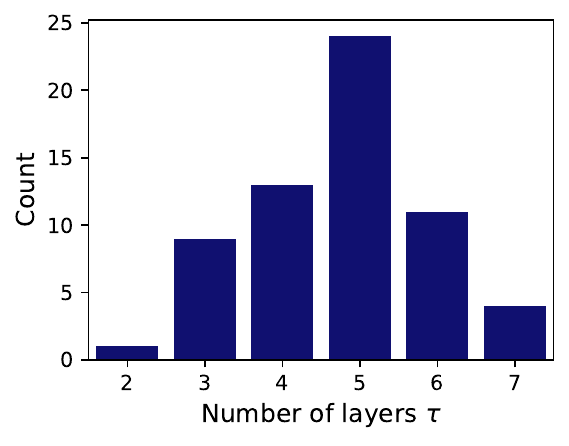} \\
         \bottomrule
    \end{tabular}
\end{table}

\begin{table}[h]
    \centering
    \caption{Distribution of the optimal number of layers at each iteration of our recursive method on the Random \#1 dataset. Each row contains egos with different numbers of total iterations, respectively 0, 1, 2, and 3.}
    \label{tab:iterations-recursive-method-random1}
    \begin{tabular}{@{}cccc@{}} 
    \toprule
    \textbf{No cut} & \textbf{1st iteration} & \textbf{2nd iteration} & \textbf{3rd iteration} \\
    \midrule
    \includegraphics[width=0.22\linewidth]{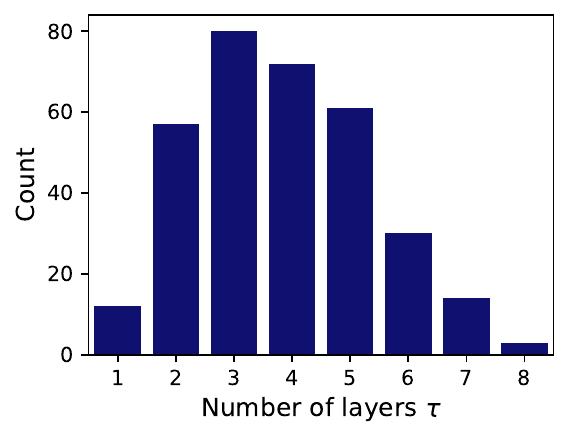} &   & & \\ \hline
        \includegraphics[width=0.22\linewidth]{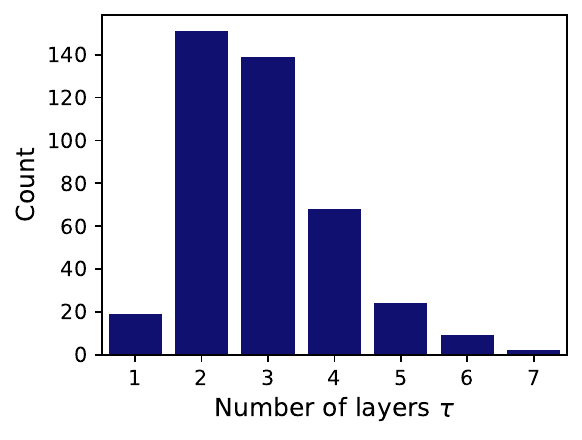} &  \includegraphics[width=0.22\linewidth]{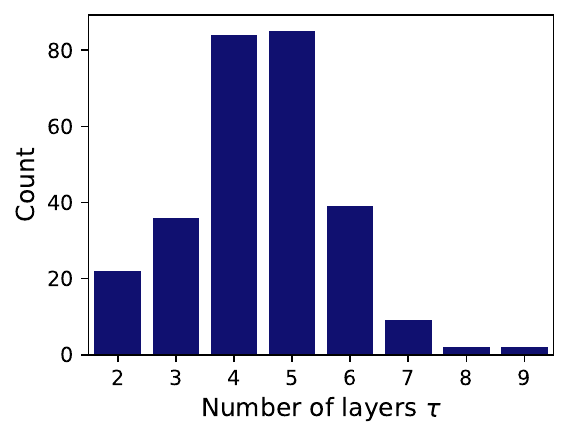} & & \\ \hline
         \includegraphics[width=0.22\linewidth]{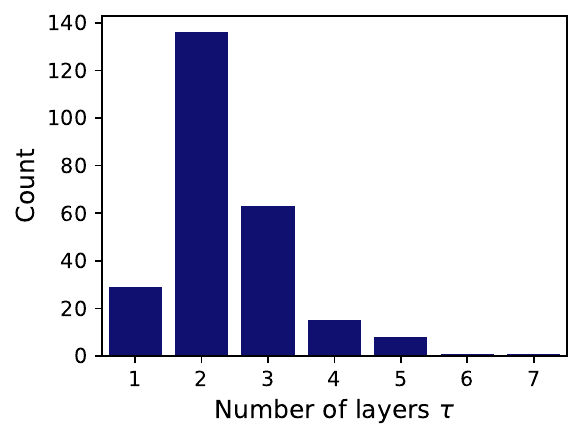} &  \includegraphics[width=0.22\linewidth]{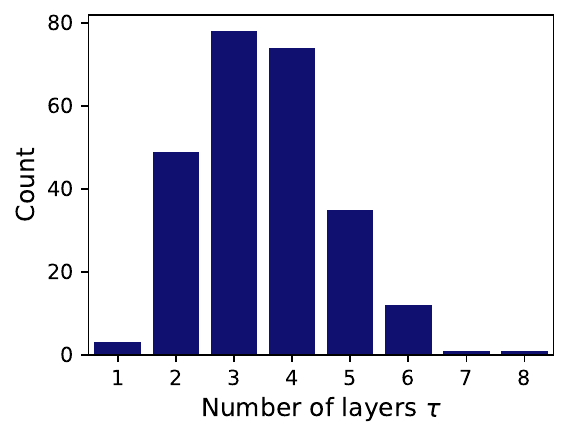} & 
         \includegraphics[width=0.22\linewidth]{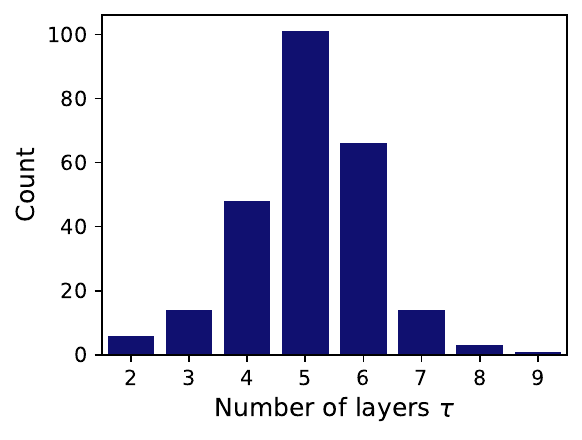} & \\ \hline
         \includegraphics[width=0.22\linewidth]{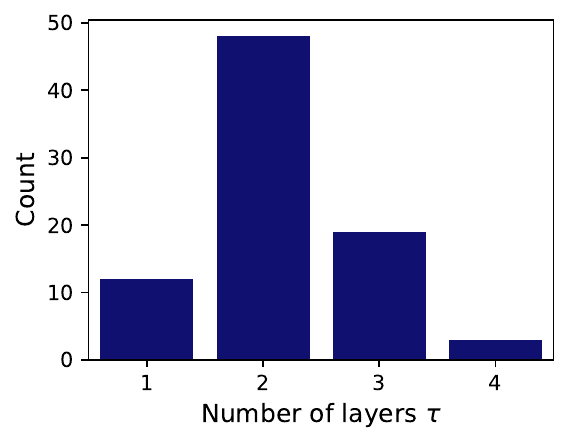} &  \includegraphics[width=0.22\linewidth]{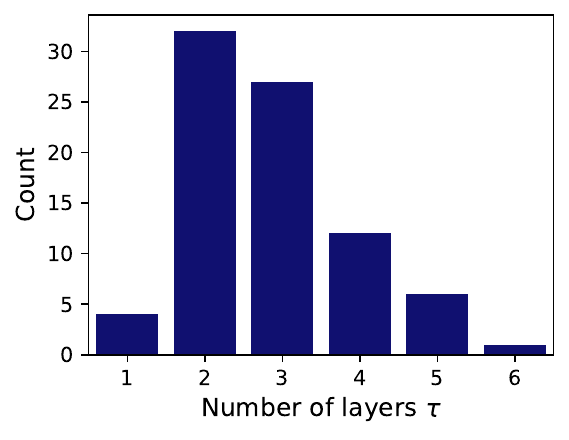} & 
         \includegraphics[width=0.22\linewidth]{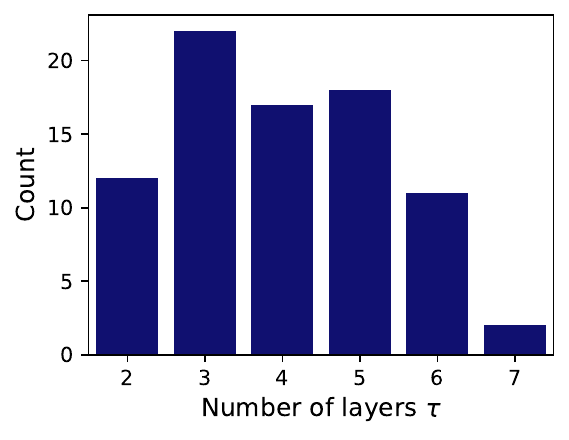} & 
         \includegraphics[width=0.22\linewidth]{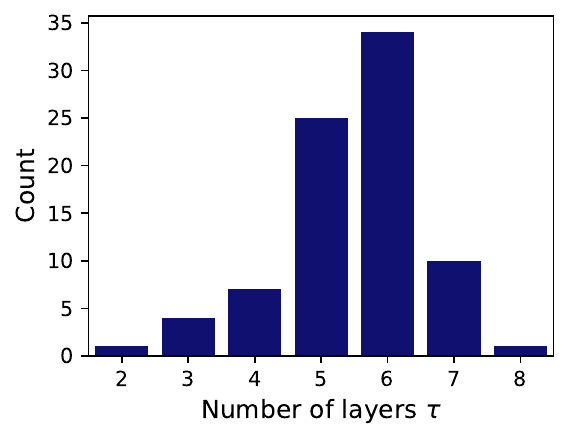}\\ 
         \bottomrule
    \end{tabular}
\end{table}

\begin{table}[h]
    \centering
    \caption{Distribution of the optimal number of layers at each iteration of our recursive method on the Random \#2 dataset. Each row contains egos with different numbers of total iterations, respectively 0, 1, 2, and 3.}
    \label{tab:iterations-recursive-method-random2}
    \begin{tabular}{@{}cccc@{}} 
    \toprule
    \textbf{No cut} & \textbf{1st iteration} & \textbf{2nd iteration} & \textbf{3rd iteration} \\
    \midrule
    \includegraphics[width=0.22\linewidth]{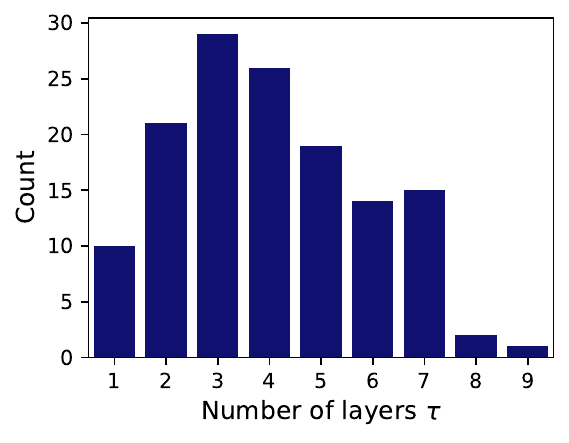} &   & & \\ \hline
        \includegraphics[width=0.22\linewidth]{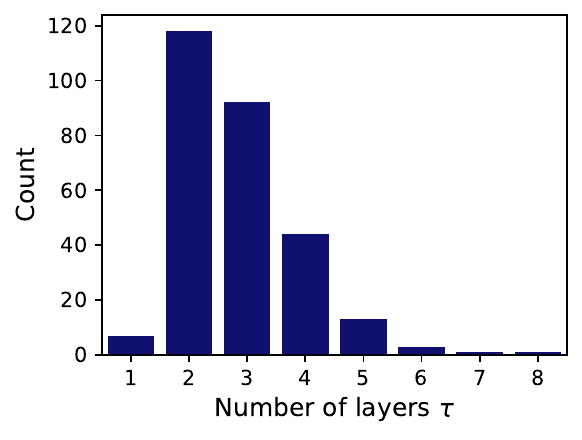} &  \includegraphics[width=0.2225\linewidth]{figures/sequential_breaks/random2/1-1.pdf} & & \\ \hline
         \includegraphics[width=0.22\linewidth]{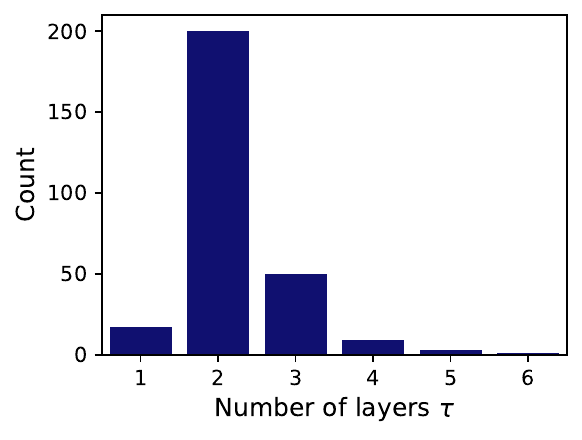} &  \includegraphics[width=0.22\linewidth]{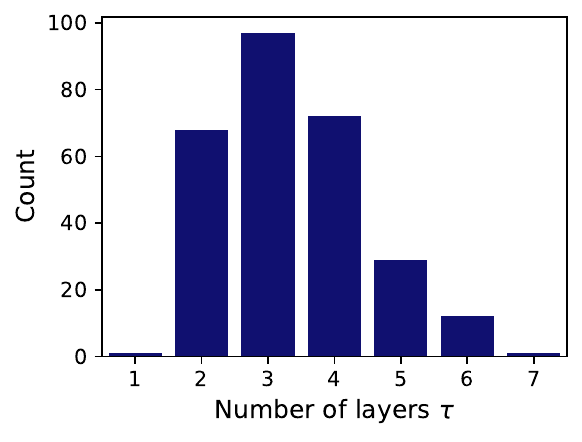} & 
         \includegraphics[width=0.22\linewidth]{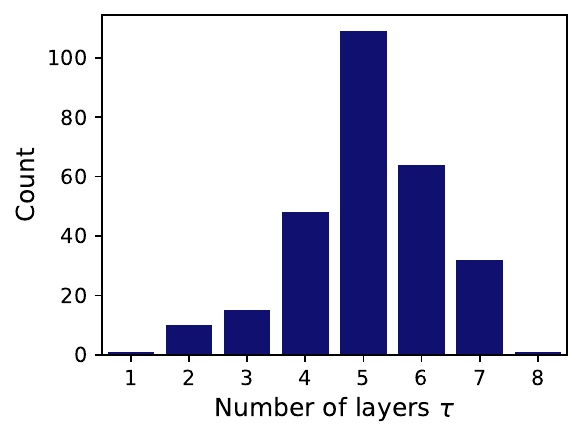} & \\ \hline
         \includegraphics[width=0.22\linewidth]{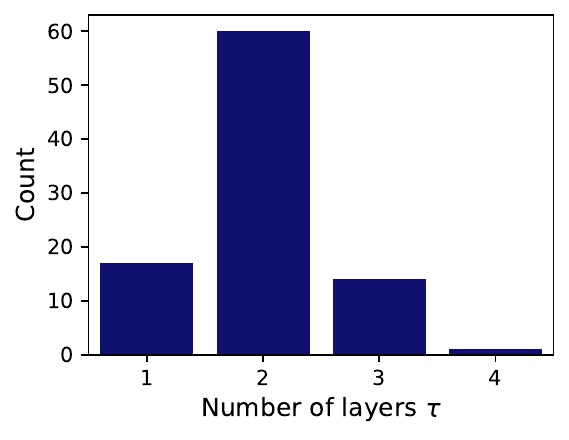} &  \includegraphics[width=0.22\linewidth]{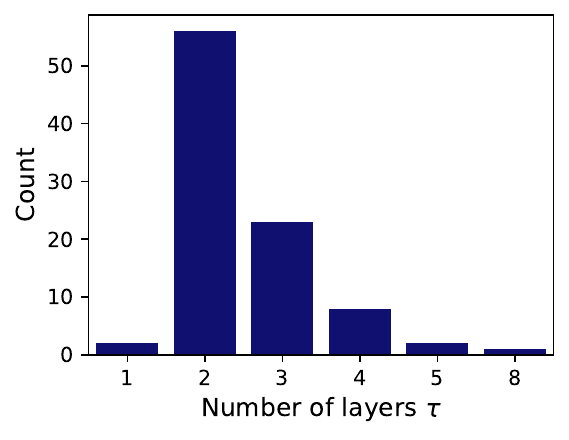} & 
         \includegraphics[width=0.22\linewidth]{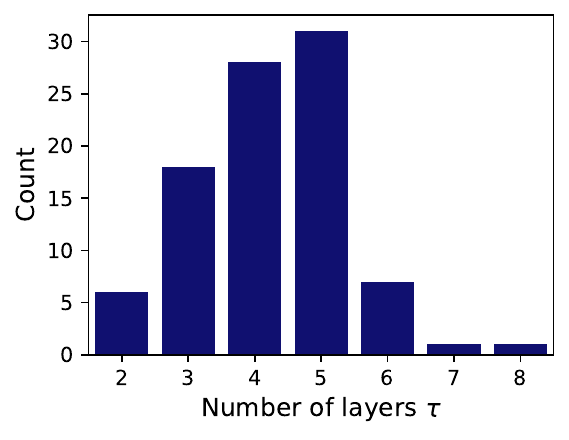} & 
         \includegraphics[width=0.22\linewidth]{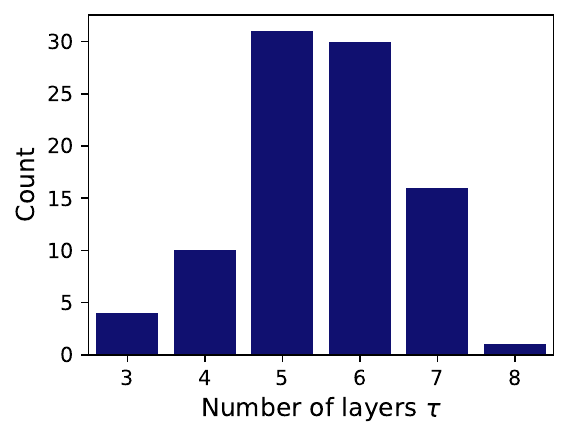}\\ 
         \bottomrule
    \end{tabular}
\end{table}